\documentclass[prd,twocolumn,preprintnumbers,nofootinbib]{revtex4-1}

\usepackage{aas_macros}
\usepackage{scrextend}
\usepackage{amsmath,bm}
\usepackage{epsfig}
\usepackage{graphicx}
\usepackage{caption}
\usepackage{subcaption}
\usepackage{graphics}
\usepackage{color}
\usepackage[normalem]{ulem}
\usepackage[breaklinks, plainpages=false, colorlinks=true, anchorcolor=cyan, linkcolor=red, citecolor=cyan, urlcolor=magenta, bookmarks=false]{hyperref}

\begin{document}
\renewcommand{\thefigure}{\arabic{figure}}
\setcounter{figure}{0}

 \def\I{{\rm i}}
 \def\E{{\rm e}}
 \def\D{{\rm d}}

\bibliographystyle{apsrev}

\title{Constraining alternative polarization states of gravitational waves from individual black hole binaries using pulsar timing arrays}

\author{Logan O'Beirne}
\affiliation{eXtreme Gravity Institute, Department of Physics, Montana State University, Bozeman, Montana 59717, USA}

\author{Neil J. Cornish}
\affiliation{eXtreme Gravity Institute, Department of Physics, Montana State University, Bozeman, Montana 59717, USA}

\author{Sarah J. Vigeland}
\affiliation{Center for Gravitation, Cosmology and Astrophysics, Department of Physics, University of Wisconsin-Milwaukee,
	P.O. Box 413, Milwaukee, WI 53201, USA}

\author{Stephen R. Taylor}
\affiliation{Theoretical AstroPhysics Including Relativity \& Cosmology (TAPIR), MC 350-17, California Institute of Technology, Pasadena, California 91125, USA}

\begin{abstract} 
Pulsar timing arrays are sensitive to gravitational wave perturbations produced by individual supermassive black hole binaries during their early inspiral phase. Modified gravity theories allow for the emission of gravitational dipole radiation, which is enhanced relative to the quadrupole contribution for low orbital velocities, making the early inspiral an ideal regime to test for the presence of modified gravity effects. Using a theory-agnostic description of modified gravity theories based on the parametrized post-Einsteinian framework, we explore the possibility of detecting deviations from General Relativity using simulated pulsar timing array data, and provide forecasts for the constraints that can be achieved. We generalize the {\tt enterprise} pulsar timing software to account for possible additional polarization states and modifications to the phase evolution, and study how accurately the parameters of simulated signals can be recovered.  We find that while a pure dipole model can partially recover a pure quadrupole signal, there is little possibility for confusion when the full model with all polarization states is used. With no signal present, and using noise levels comparable to those seen in contemporary arrays, we produce forecasts for the upper limits that can be placed on the amplitudes of alternative polarization modes as a function of the sky location of the source.
\end{abstract}

\maketitle

\section{Introduction}
The dark energy and dark matter problems in cosmology and the unresolved reconciliation between General Relativity (GR) and quantum mechanics suggest that Einstein's theory of gravity is incomplete~\cite{beyondeinstein}. Gravitational wave (GW) astronomy provides a new arena to search for deviations from GR. One smoking gun signature would be the detection of additional polarization states. Many theories that violate the strong equivalence principle or Lorentz invariance allow for the emission of dipole radiation~\cite{1977ApJ...214..826W,Chatziioannou:2012rf,2016PhRvD..94h4002Y}. To search for this signature it is best to observe binary systems that are widely separated since deviations from pure quadrupole emission are enhanced for low velocity systems~\cite{Chatziioannou:2012rf}. It is also advantageous to measure multiple independent projections of the polarization pattern~\cite{Chatziioannou:2012rf}. 

The GW detections that have been made by the LIGO-Virgo instruments are of the high velocity final inspiral and merger phase, where there is only a small amplification of the dipole/tensor ratio. Moreover,  there are presently a limited number of ground-based detectors, providing a limited number of projections of the radiation field, so it difficult to decipher the polarization pattern~\cite{Abbott:2017tlp,Abbott:2017oio,Abbott:2018utx,Chatziioannou:2012rf}. Pulsar timing arrays are well suited to constraining dipole emission since they observe in a frequency band where supermassive black hole binaries will be moving relatively slowly, and with dozens of pulsars in the array, they provide multiple projections of the radiation field.

Rapidly rotating neutron stars, known as pulsars, emit beams of electromagnetic radiation that are observed as radio pulses when the beam sweeps across the Earth. Millisecond pulsars that have been spun up due to accretion act as very stable clocks whose pulse phases are known to high precision~\cite{2010CQGra..27h4013H}, allowing astronomers to search for slight perturbations in the times of arrival (TOA) of radio pulses caused by low frequency GWs~\cite{1975GReGr...6..439E,1979ApJ...234.1100D}. A collection of these comprise a pulsar timing array (PTA), a galactic scale GW detector. There are currently three distinct PTAs operating around the world~\cite{2016MNRAS.458.3341D,2013CQGra..30v4008M,2013CQGra..30v4007H} whose combined efforts comprise the International Pulsar Timing Array~\cite{Verbiest:2016vem}.

PTAs observe frequencies of approximately $10^{-9}-10^{-7}$ Hz, and it is believed that the dominant source of GWs in this frequency band is produced by a population of supermassive black hole binaries (SMBHBs) in their slow, adiabatic inspiral phase~\cite{2003ApJ...583..616J,2008MNRAS.390..192S,2009MNRAS.394.2255S}. Modeling suggests that the ensemble signal from multiple binary systems will be detect first, followed by the signals from the loudest individual systems~\cite{Rosado:2015epa}. PTAs probe a regime well before SMBHBs merge, where the systems have orbital velocities of order $v/c \sim 10^{-2}-10^{-1}$, and where any dipole radiation will be enhanced by a factor of $10 - 100$ relative to the quadrupole.

Here we study how the signals from individual SMBHBs can be used to constrain alternative theories of gravity. We use the model independent parametrized post-Einsteinian formulation~\cite{Yunes:2009ke,Chatziioannou:2012rf} of modified gravitational theories to model simulated signals from individual SMBHBs that include all polarizations allowed by a general metric theory of gravity. We then use Bayesian inference to study the signals from simulated pulsar timing data sets. We explore how well the system parameters can be recovered, and the upper limits that can be placed when no signal is present in the data. 

 In Section~\ref{section2}, we describe the post-Einsteinian signal model and make comparisons with the GR model. Section~\ref{section3} outlines the data generation and analysis methods. Section~\ref{sectionresults} explores how well the model parameters can be recovered from simulated data, and in the absence of a signal, how the upper limits on the amplitudes of each polarization mode will depend on sky location. In Section~\ref{conclusions} we present our conclusions. Throughout this paper, we use units where $G=c=1$.

\section{Signal model}\label{section2}
Pulsar timing arrays encode GWs in the timing residuals, which are found by subtracting the timing model from the raw arrival times\footnote{The timing model includes relativistic effects such as the Shapiro time delay and Einstein delay, and these effects are modified in alternative theories of gravity. However, existing solar system constraints on these post-Newtomian effects, which scale with the PPN parameter $\gamma$, are of order pico-seconds~\cite{Sampson:2013wia},  so we can safely use the standard GR timing model in our analysis}. A single pulsar's timing residual $\delta t$ can be written:
\begin{eqnarray}\label{sigmodel}
\delta t = \bm{M}\cdot \delta\bm{\xi} + n + s,
\end{eqnarray}
where $\bm{M}\cdot \delta\bm{\xi}$ describes uncertainties in the timing model~\cite{Ellis_2013,2013MNRAS.428.1147V,2013CQGra..30v4004E,2012ApJ...756..175E}, $n$ is the white noise, and $s$ is the GW signal. We omit red noise in our simulations as including it in the noise model significantly slows down the likelihood evaluations. Leaving out the red noise results in somewhat optimistic predictions for the signal extraction capabilities and the strength of the upper limits.

For a gravitational wave propagating in the $\bm{\Omega}$ direction we can introduce the orthonormal coordinate system
\begin{eqnarray}
&& \bm{\Omega} \rightarrow (-\text{sin}\,\theta\,\text{cos}\,\phi,-\text{sin}\,\theta\,\text{sin}\,\phi,-\text{cos}\,\theta) \nonumber \\
&& \textbf{u} \rightarrow (\text{sin}\,\phi,-\text{cos}\,\phi,0) \nonumber \\
&& \textbf{v} \rightarrow (\text{cos}\,\theta\,\text{cos}\,\phi,\text{cos}\,\theta\,\text{sin}\,\phi,-\text{sin}\,\theta)\, ,
\end{eqnarray}
These are related to the principle axes $\bf{m}$, $\bf{n}$ of the source by a rotation angle $\psi$ (the polarization angle) about the propagation direction:
\begin{eqnarray}
&& \bf{m} = \bf{u}\,\text{cos}\,\psi + \bf{v}\,\text{sin}\,\psi \nonumber \\
&& \bf{n} = -\bf{u}\,\text{sin}\,\psi + \bf{v}\,\text{cos}\,\psi\, ,
\end{eqnarray}
The basis tensors for the various gravitational wave polarization states are then:
\begin{eqnarray}
&& \bm{\epsilon}_{\text{TT}}^{+} = \textbf{u}\otimes\textbf{u} - \textbf{v}\otimes\textbf{v} = \text{cos}(2\psi)\textbf{e}^{+} - \text{sin}(2\psi)\textbf{e}^{\times} \nonumber \\
&& \bm{\epsilon}_{\text{TT}}^{\times} = \textbf{u}\otimes\textbf{v} + \textbf{v}\otimes\textbf{u} = \text{sin}(2\psi)\textbf{e}^{+} + \text{cos}(2\psi)\textbf{e}^{\times}  \nonumber \\
&& \bm{\epsilon}_{\text{ST}}^{\odot} = \textbf{u}\otimes\textbf{u} + \textbf{v}\otimes\textbf{v} =\textbf{m}\otimes\textbf{m} + \textbf{n}\otimes\textbf{n} \nonumber \\
&& \bm{\epsilon}_{\text{VL}}^{\text{u}} = \textbf{u}\otimes\bm{\Omega} + \bm{\Omega}\otimes\textbf{u} = \text{cos}(\psi)\,\textbf{e}^{\text{u}} - \text{sin}(\psi)\,\textbf{e}^{\text{v}} \nonumber \\
&& \bm{\epsilon}_{\text{VL}}^{\text{v}} = \textbf{v}\otimes\bm{\Omega} + \bm{\Omega}\otimes\textbf{v} = \text{sin}(\psi)\,\textbf{e}^{\text{u}} + \text{cos}(\psi)\,\textbf{e}^{\text{v}} \nonumber \\
&& \bm{\epsilon}_{\text{SL}}^{\leftrightarrow} = \bm{\Omega}\otimes\bm{\Omega} ,
\end{eqnarray}
where
\begin{eqnarray}
&& \textbf{e}^{+} = \textbf{m}\otimes\textbf{m} - \textbf{n}\otimes\textbf{n} \nonumber \\
&& \textbf{e}^{\times} = \textbf{m}\otimes\textbf{n} + \textbf{n}\otimes\textbf{m}  \nonumber \\
&& \textbf{e}^{\odot} = \bm{\epsilon}^{\odot}_{\text{ST}}  \nonumber \\
&& \textbf{e}^{\text{u}} = \textbf{m}\otimes\bm{\Omega} + \bm{\Omega}\otimes\textbf{m} \nonumber \\
&& \textbf{e}^{\text{v}} = \textbf{n}\otimes\bm{\Omega} + \bm{\Omega}\otimes\textbf{n} \nonumber \\
&& \textbf{e}^{\leftrightarrow} = \bm{\epsilon}^{\leftrightarrow}_{\text{SL}} ,
\end{eqnarray}
and the subindices are labeled for the 2 tensor transverse (TT) modes of General Relativity ($+$ and $\times$), a scalar transverse (ST) ``breathing'' mode ($\odot$), 2 vector longitudinal (VL) modes ($u$ and $v$), and a scalar longitudinal (SL) mode ($\leftrightarrow$) . We refer to the latter four as alternative polarizations (alt-pols). The timing residuals induced by polarization state \textit{A} for a pulsar in the \textbf{p} direction is given by
\begin{eqnarray}\label{residual}
&& r^{A}(t_{e}) = \frac{1}{2(1 + \bm{\Omega}\cdot\textbf{p})}\textbf{p}\otimes\textbf{p}:(\textbf{H}^{A}(t_{e})-\textbf{H}^{A}(t_{p})) ,
\end{eqnarray}
where $t_{p} = t_{e} - L(1 + \bm{\Omega}\cdot\textbf{p})$, $t_{e}$ is the time at Earth, $L$ is the distance to the pulsar and $\textbf{H}^{A} = \int^{t}\textbf{h}^{A}dt$ is the anti-derivative of the gravitational wave strain. Note that there are contributions from the presence of GWs at the pulsar and the Earth, indicated in Eq.(\ref{residual}). We assume all modes travel at the speed of light. However, Lorentz-violating and massive gravity allow for superluminal~\cite{2006CQGra..23.5643E,2007PhRvD..76h4033F,2011PhRvD..83l4043B,2016PhRvD..93d4044B,Yagi:2013ava,2015PhRvD..91h2003H} and subluminal propagation~\cite{2016PTEP.2016j3E02K,2015CQGra..32o4001B,2015LNP...892..161V,beyondeinstein,lrr-2013-9} of non-Einsteinian modes, respectively. Superluminal modes decrease the effective luminosity distance to the binary and the pulsar frequency in that respective alt-pol's response. This would be an interesting extension to the current study, but we do not anticipate that the upper limits would be change significantly in this case as it only impacts the pulsar terms, which are less constraining than the Earth term due to uncertainties in the pulsar distances. In the case of massive gravity, the resulting dispersion relationship makes the analysis appreciably more complicated and is beyond the scope of this study.

\begin{figure*}
	\captionsetup[subfigure]{labelformat=empty}
	\centering
	\begin{subfigure}[c]{0.03\textwidth}
	    TT/\\ST
	\end{subfigure}
	~
	\begin{subfigure}[c]{0.3\textwidth}
		\includegraphics[width=\textwidth]{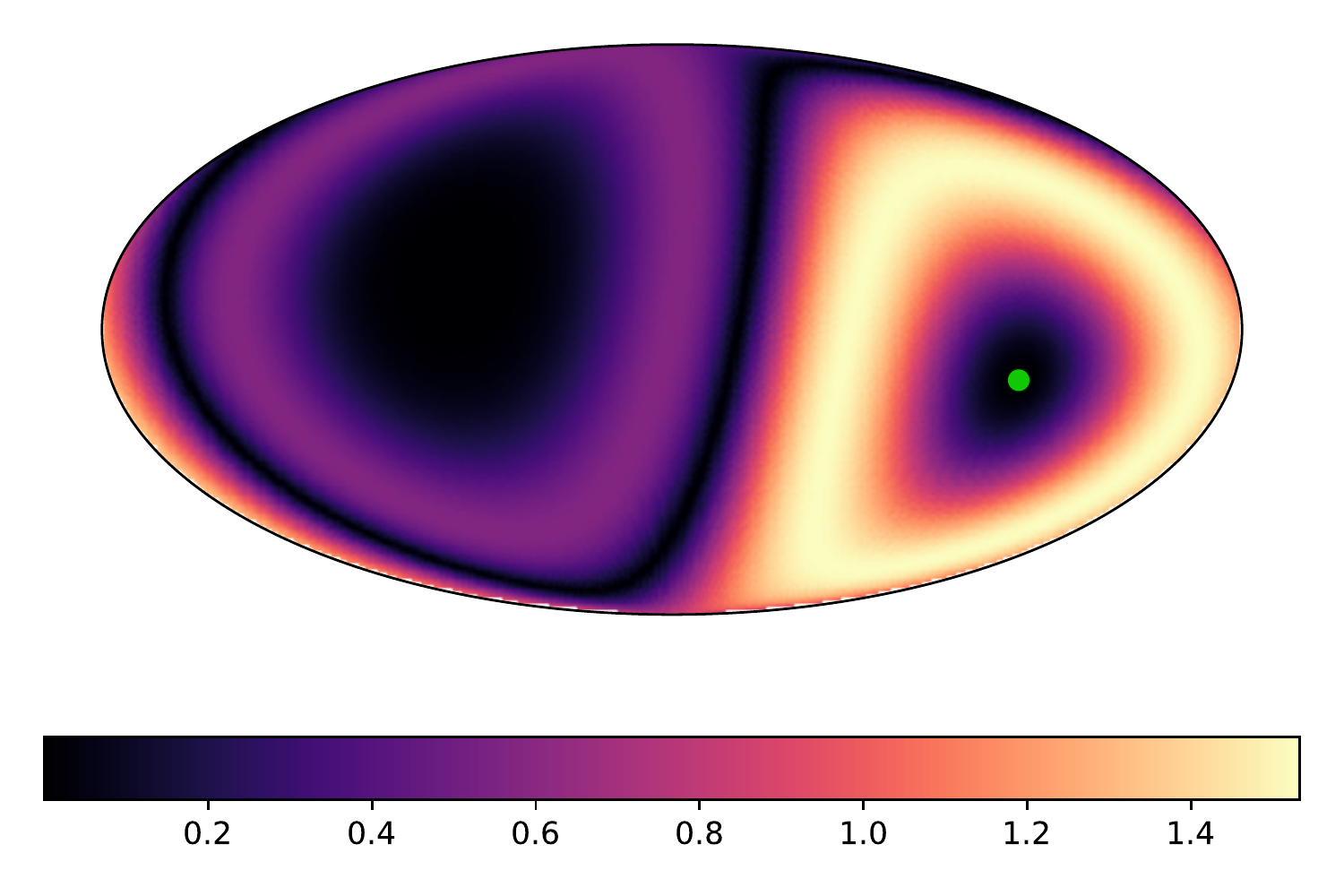}
	\end{subfigure}
	~ 
	\begin{subfigure}[c]{0.3\textwidth}
		\includegraphics[width=\textwidth]{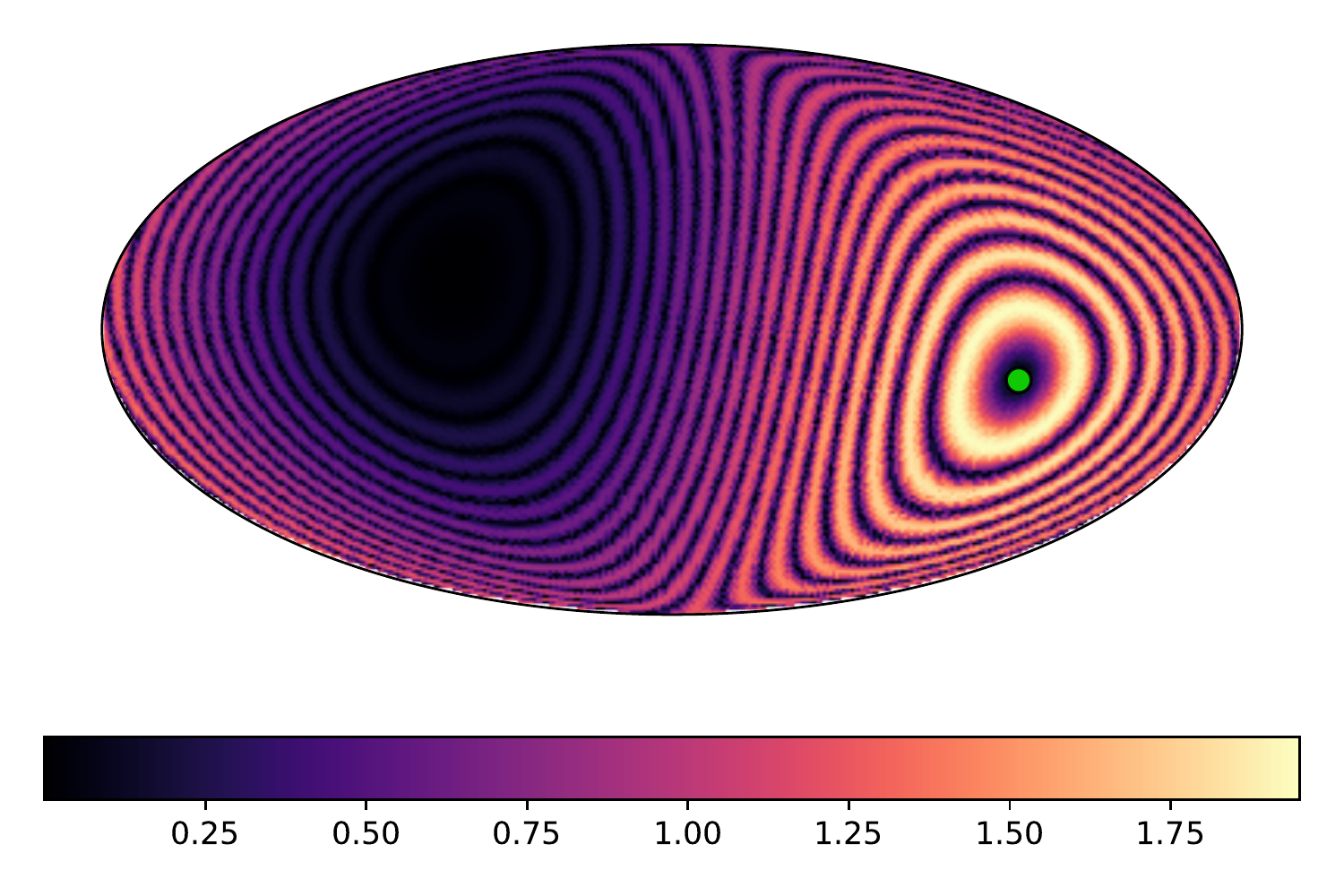}
	\end{subfigure}
	~ 
	\begin{subfigure}[c]{0.3\textwidth}
		\includegraphics[width=\textwidth]{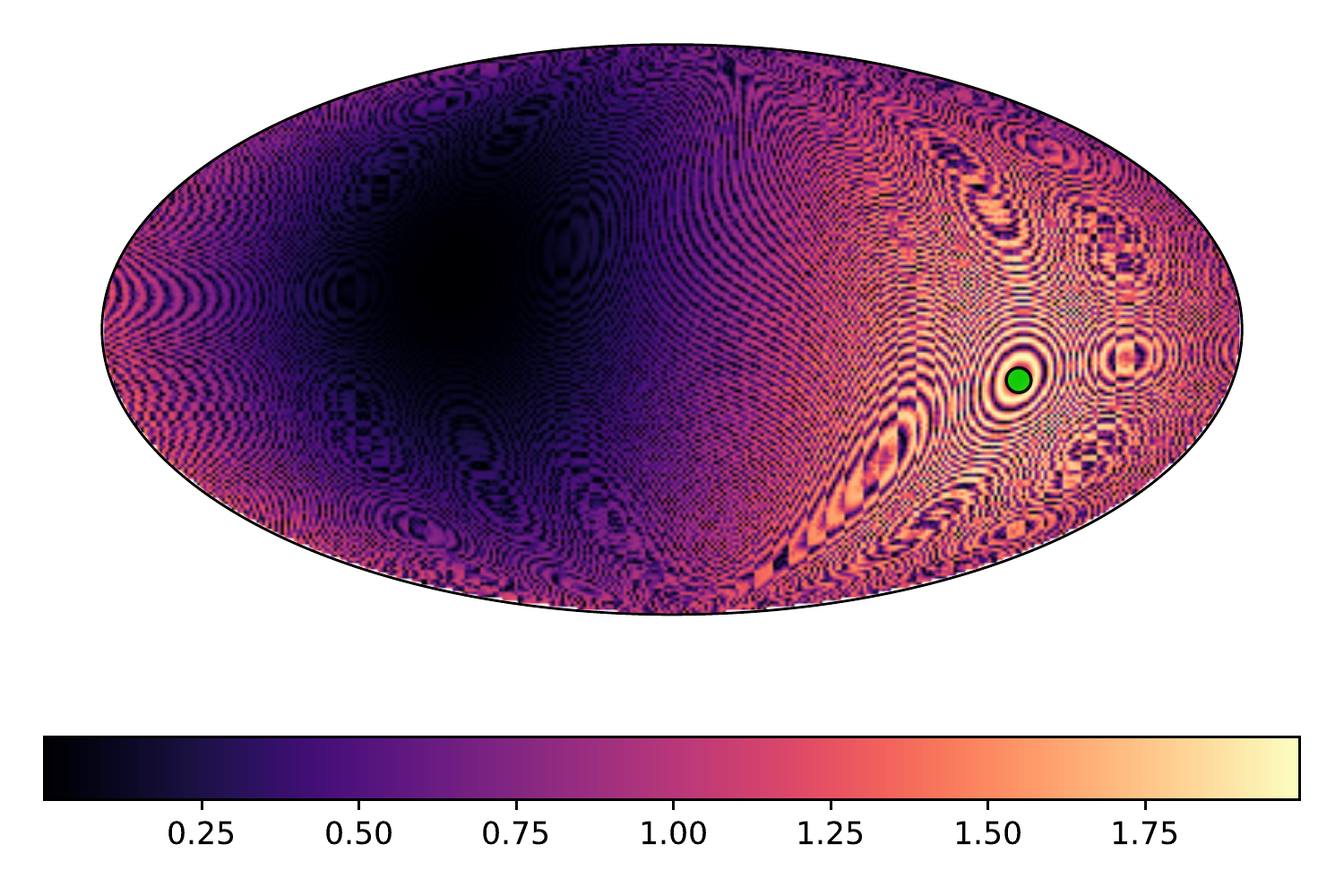}
	\end{subfigure}
	
	\begin{subfigure}[c]{0.03\textwidth}
		VL 
	\end{subfigure}
	\begin{subfigure}[c]{0.3\textwidth}
		\includegraphics[width=\textwidth]{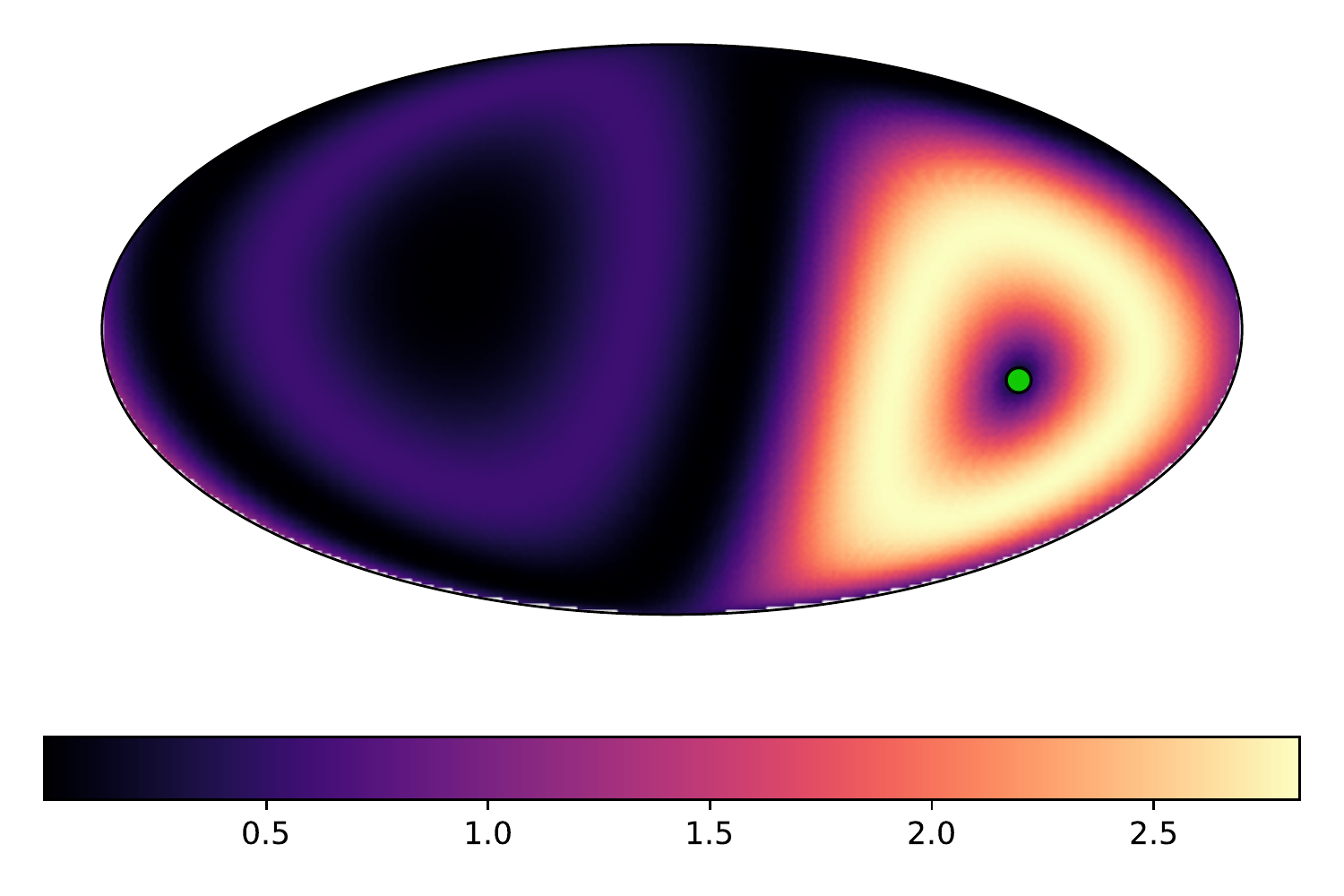}
	\end{subfigure}
	~ 
	\begin{subfigure}[c]{0.3\textwidth}
		\includegraphics[width=\textwidth]{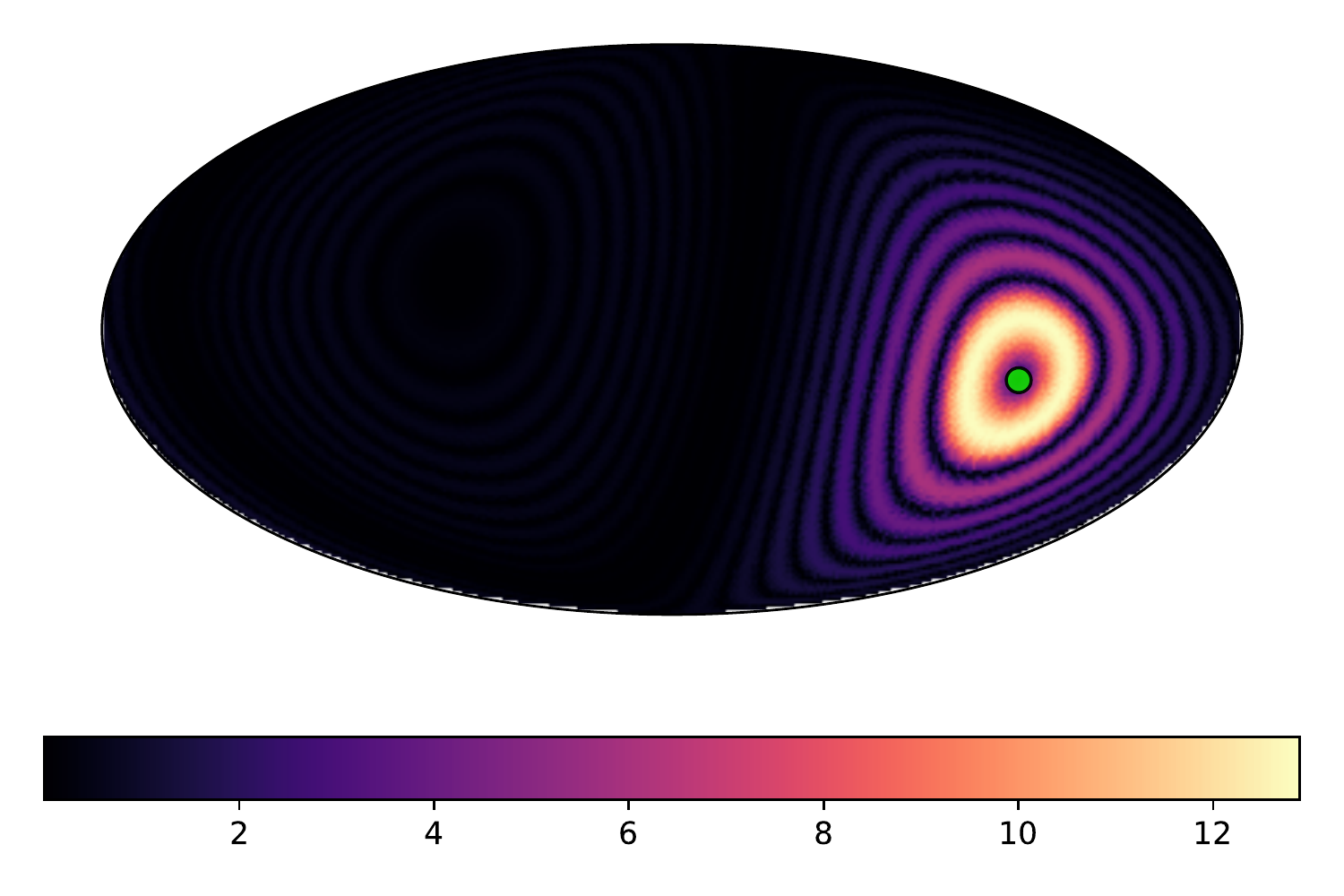}
	\end{subfigure}
	~ 
	\begin{subfigure}[c]{0.3\textwidth}
		\includegraphics[width=\textwidth]{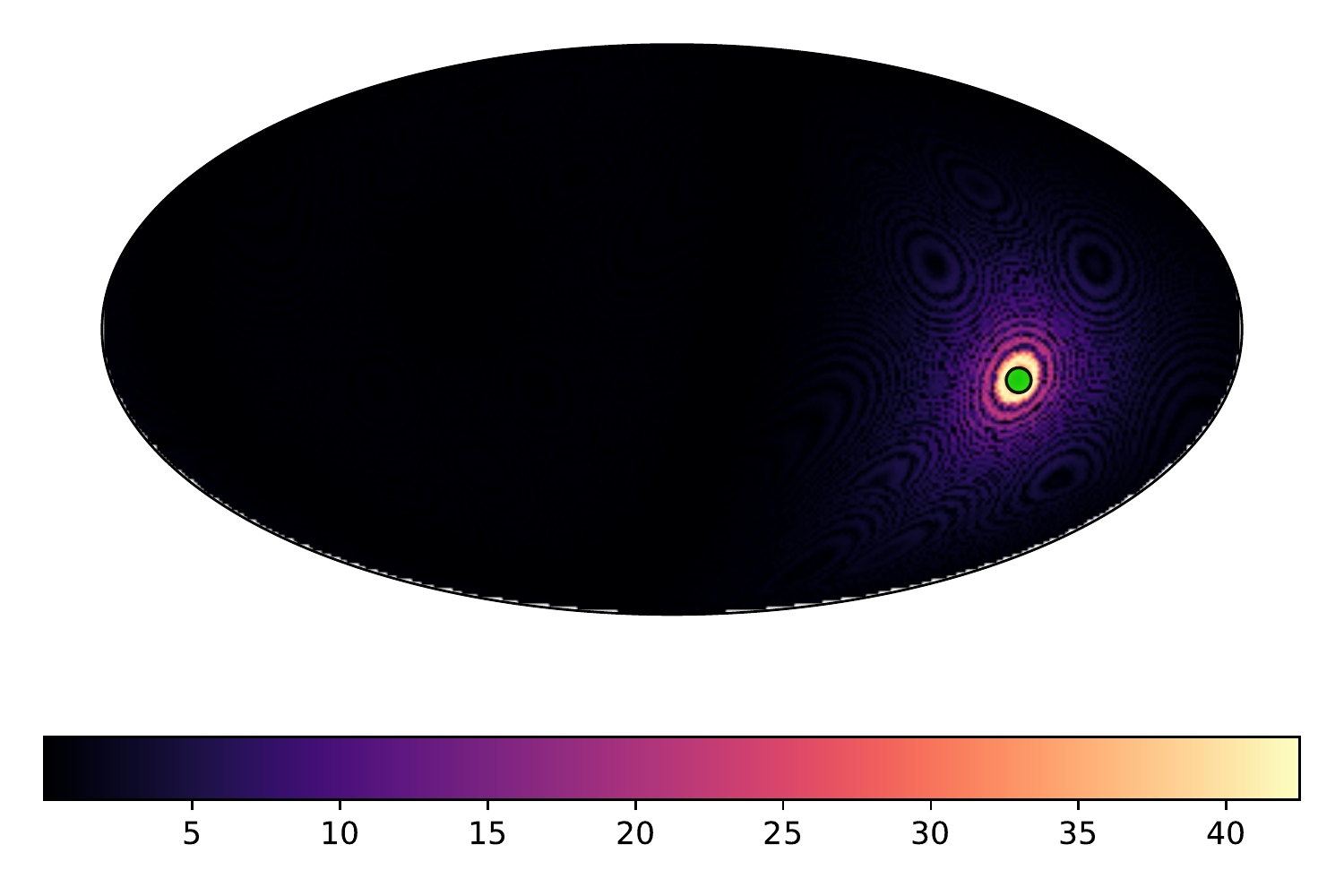}
	\end{subfigure}
	
	\begin{subfigure}[c]{0.03\textwidth}
		SL 
	\end{subfigure}
	\begin{subfigure}[c]{0.3\textwidth}
		\includegraphics[width=\textwidth]{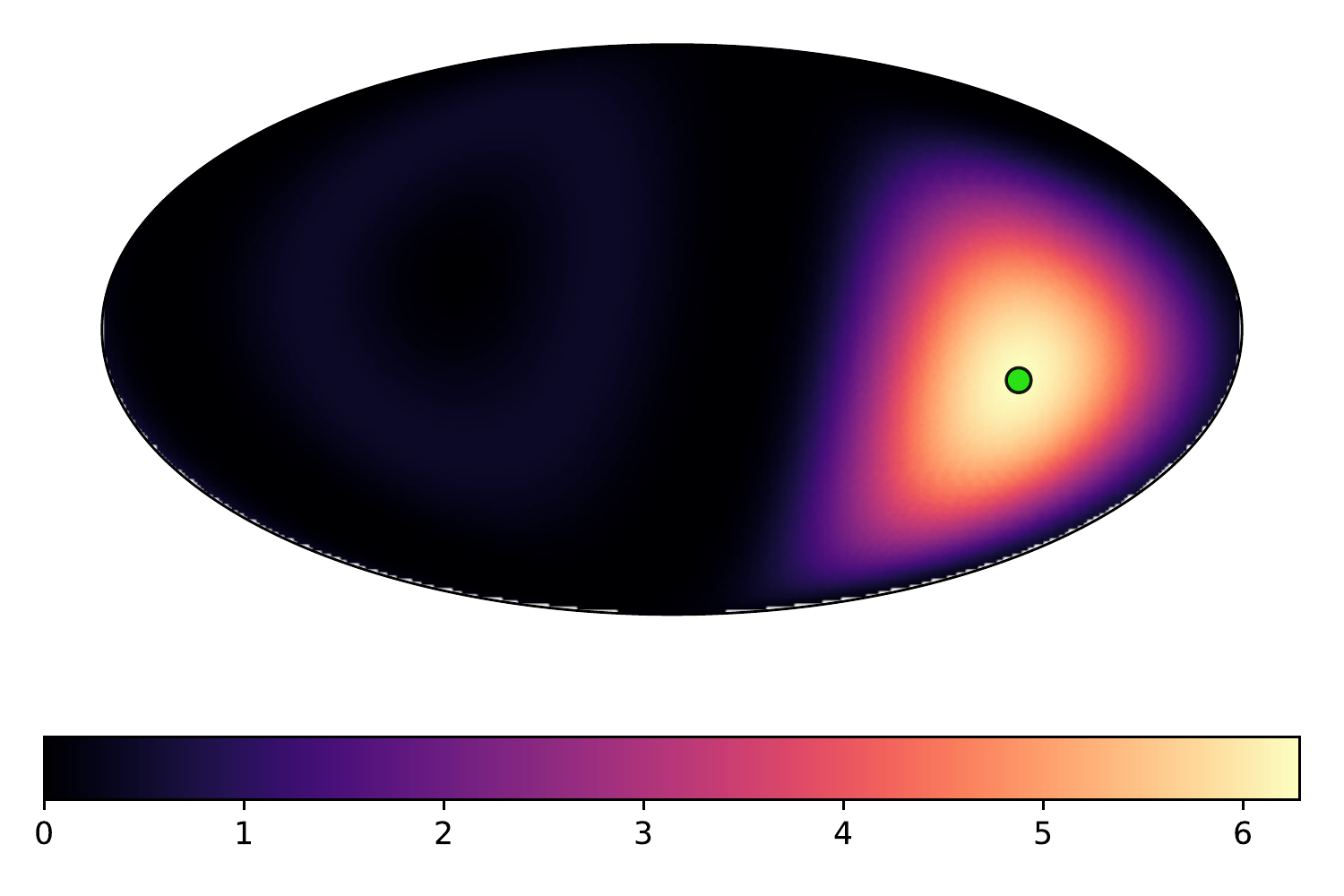}
		\caption{$fL=1$}
	\end{subfigure}
	~ 
	\begin{subfigure}[c]{0.3\textwidth}
		\includegraphics[width=\textwidth]{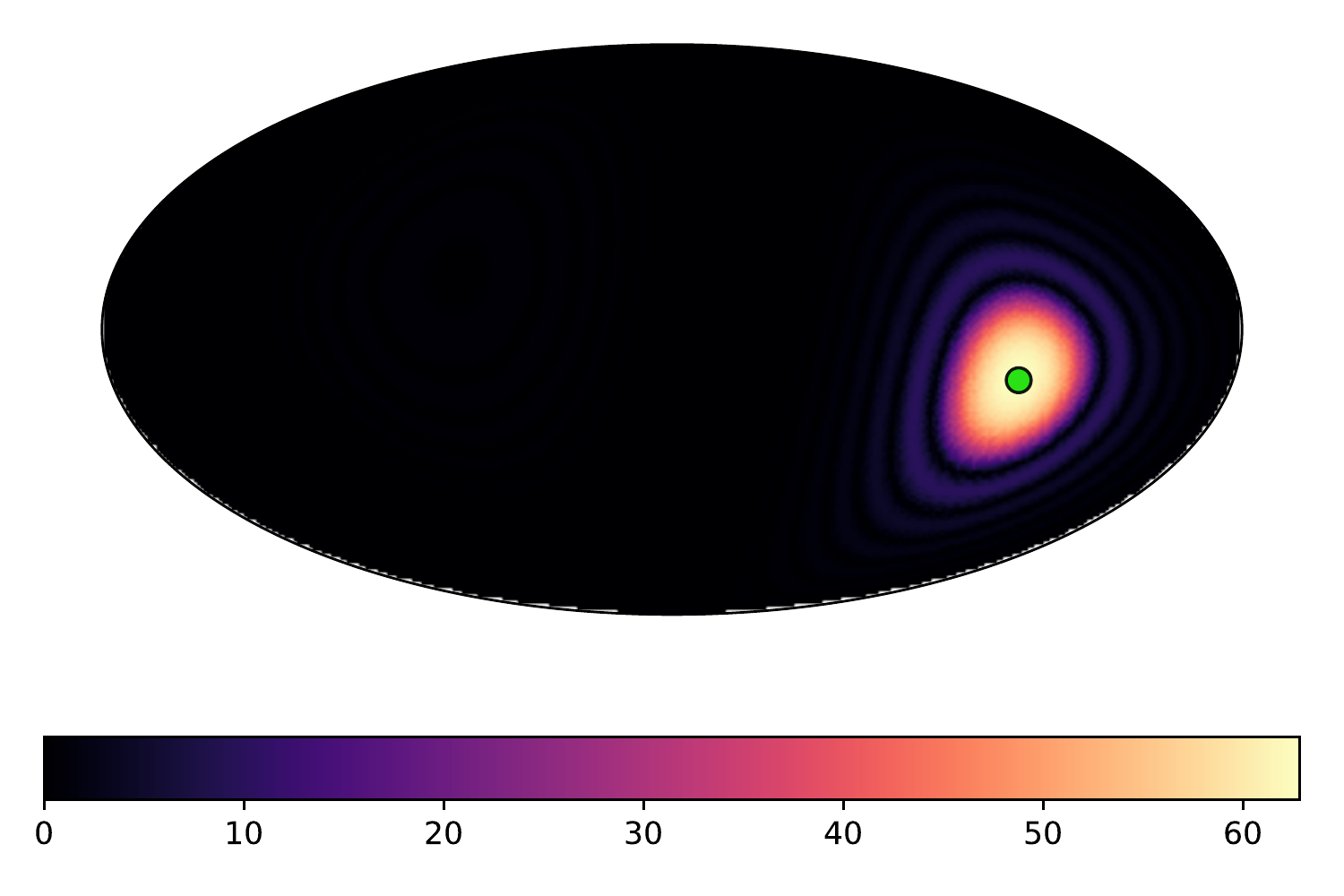}
		\caption{$fL=10$}
	\end{subfigure}
	~ 
	\begin{subfigure}[c]{0.3\textwidth}
		\includegraphics[width=\textwidth]{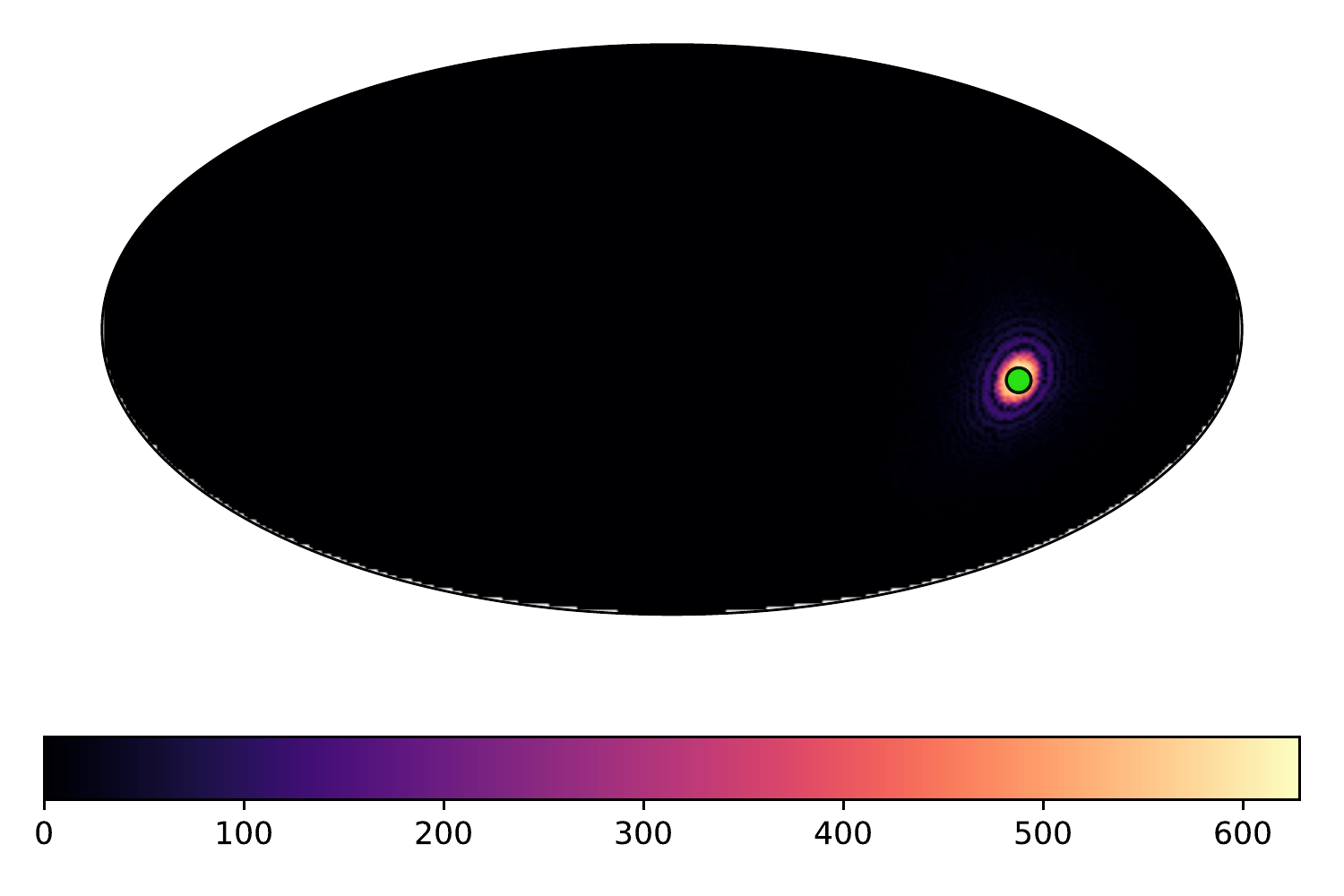}
		\caption{$fL=100$}
	\end{subfigure}
	\caption{GW antenna pattern sky maps. The color gradient indicates how sensitive the response is to a GW based on the pulsar's sky position. The GW is originating from green circle. From left to right, the columns correspond to values of fL = 1, 10, and 100, respectively. From top to bottom, the polarizations are the rms of TT/ST, VL, and SL, respectively.}
	\label{fig:antenna}
\end{figure*}

We define the frequency dependent antenna patterns as
\begin{eqnarray}\label{Fpatteq}
&& F^{A} = \left[ \frac{|e^{-2\pi i f L(1 + \bm{\Omega}\cdot\textbf{p})}-1|}{2(1 + \bm{\Omega}\cdot\textbf{p})}\right] \,\left( \textbf{e}^{A}:\textbf{p}\otimes\textbf{p} \right).
\end{eqnarray}
The frequency dependent prefactor in square brackets oscillates rapidly for large $fL$, and Ref.~\cite{2012ApJ...756..175E} argued that the oscillating term should be dropped. Here we are required to include this factor to avoid singularities in the longitudinal antenna patterns when the pulsar is in the same direction as the source: $1+\bm{\Omega}\cdot\textbf{p}=0$. Furthermore, as discussed in Ref.~\cite{2010arXiv1008.1782C}, its inclusion improves sky localization. Figure~\ref{fig:antenna} shows the sky maps of various polarization antenna patterns at different values of $fL$ -- the number of gravitational wavelengths to the pulsar.
For the longitudinal modes there is increased sensitivity in the direction of propagation~\cite{2012PhRvD..85h2001C,2011PhRvD..83l3529A}, proportional to $\delta fL$ for the VL mode and $fL$ for the SL mode where $\delta$ is the small angle subtended by the pulsar and GW propagation direction. The transverse modes are not significantly enhanced since the response scales as $\delta^{2} fL$. The enhanced response for pulsars near the line of sight to the GW source increases with increasing $fL$, as can be seen going from left to right along the lower two rows in Figure~\ref{fig:antenna}. It is worth noting that the transverse modes excite zero response in pulsars in exactly the same direction as the source, or in the antipodal direction, while  the SL mode excites no response in pulsars that are oriented perpendicular to the line of sight to the source. Similarly, the VL modes excite no response in pulsars in towards or antipodal to the source, or in those perpendicular to the line of sight. All modes are less sensitive when the pulsar is the opposite direction to the GW source. To reiterate, the GW response to all polarization modes is largest when the pulsar is in almost the same sky direction as the source. The longitudinal modes have an enhanced response relative to the transverse modes in this respect, but the transverse modes have better sky coverage. Depending on where the GW source is located with respect to pulsars in the array, there is potential to have tighter constraints on the strains of longitudinal modes, or for the longitudinal modes to be completely undetectable to the array. 

For the GW residuals in GR we have~\cite{2010arXiv1008.1782C}
\begin{eqnarray}
&& \textbf{H}^{+} = \textbf{e}^{+} \frac{\mathcal{M}^{5/3}}{d_{L}\omega^{1/3}}(1 + \text{cos}^{2}\iota)\,\text{cos}\Big( 2 \int^{t} \omega dt + 2 \Phi_{0} \Big) \nonumber \\
&&\hspace*{2.0in}\times(1 + \mathcal{O}(\dot{\omega}/\omega^{2})), \nonumber \\
&& \textbf{H}^{\times} = \textbf{e}^{\times} \frac{2\mathcal{M}^{5/3}}{d_{L}\omega^{1/3}}\text{cos}\,\iota\,\text{sin}\Big(2\int^{t} \omega dt + 2 \Phi_{0} \Big) \nonumber \\
&&\hspace*{2.0in}\times(1 + \mathcal{O}(\dot{\omega}/\omega^{2})) ,
\end{eqnarray}
where $d_{L}$ is the luminosity distance, $\mathcal{M}$ is the chirp mass, $\Phi_{0}$ is the initial orbital phase of the binary, $\omega$ is the orbital angular frequency, and $\iota$ is the angle of inclination of the binary.

Note that the expression for the anti-derivative of the gravitational wave amplitude assumes that we are working in the slow-evolution limit. We can evaluate the size of the errors that this introduces:
\begin{eqnarray}
&& \textbf{H}(t) = \int^{t}h(t)dt = \int^{\Phi(t)} \frac{A(\Phi)}{\omega(\Phi)}\text{sin}\,{\Phi}d\Phi ,
\end{eqnarray}
so that 
\begin{eqnarray}
&& \textbf{H}(t) = \frac{A(\Phi)}{\omega(\Phi)}\text{cos}\,\Phi + \Bigg(\frac{A(\Phi)}{\omega(\Phi)} \Bigg)' \text{sin}\,\Phi  \nonumber \\
&&\hspace*{1.4in}- \Bigg(\frac{A(\Phi)}{\omega(\Phi)} \Bigg)'' \text{cos}\,\Phi + ...\,\, ,
\end{eqnarray}
where the primes denote derivatives with respect to $\Phi$. Note that $d/d\Phi$ = $\omega^{-1}d/dt $. The ratio of the second order term to the first order term is given by
\begin{eqnarray}
&&\Big[ \text{ln}\Big(\frac{A_{p}}{\omega_{p}}\Big) \Big]' = - \frac{\dot{\omega}}{3\,\omega^{2}} .
\end{eqnarray}
In GR this is given to leading PN order by 
\begin{eqnarray}
&&\Big[ \text{ln}\Big(\frac{A_{p}}{\omega_{p}}\Big) \Big]' = - \frac{32}{5}\mathcal{M}^{5/3}\omega^{5/3} \nonumber \\
&& = -3.8 \times 10^{-2} \Big( \frac{\mathcal{M}}{10^{10}M_{\odot}} \Big)^{5/3}  \Big( \frac{f_{\text{GW}}}{3 \times 10^{-7} \text{Hz}}  \Big)^{5/3},
\end{eqnarray}
which validates the dropping of higher order corrections in PTA analyses.

In the current NANOGrav analysis for continuous wave sources~\cite{NG11yrCW}, upper limits are quoted on $h_{\text{TT}}$ as a function of the TT-mode frequency $2\omega_0$, where $\omega_{0}$ is the orbital frequency as measured at the Earth. In producing the upper limits the signal model is marginalized over
the GW parameters
\begin{eqnarray}
&& \vec{\lambda} \rightarrow (\theta,\phi,\Phi_{0},\psi,\iota,\mathcal{M},h_{\text{TT}})
\end{eqnarray}
and the pulsar distances $L_{i}$. To allow for alternative theories of gravity, we need to enlarge the parameter set to
\begin{eqnarray}\label{altpolparam}
&& \vec{\lambda} \rightarrow (\theta,\phi,\Phi_{0},\psi,\iota,\alpha_{D},\alpha_{Q},h_{\text{TT}},h_{\text{ST}},h_{\text{VL}},h_{\text{SL}})
\end{eqnarray}
where $h_{\text{ST}},h_{\text{VL}},h_{\text{SL}}$ are the amplitudes of the additional polarization modes, and the parameter $\alpha_{D}$, $\alpha_{Q}$ scale the dipole and quadrupole contribution to the frequency evolution:
\begin{eqnarray}
&& \frac{d\omega}{dt} = \alpha_{D}\,\omega^{3} + \alpha_{Q}\,\omega^{11/3}
\end{eqnarray}
We have neglected higher order terms in the frequency evolution since they are negligible for slowly moving sources. The GR limit is recovered by setting $\alpha_{D} = 0$ and $\alpha_{Q} = \frac{96}{5}\mathcal{M}^{5/3}$. The wave tensors are given by~\cite{Chatziioannou:2012rf,2015PhRvD..91h2003H}
\begin{eqnarray}\label{altpolresiduals}
&& \textbf{H}^{+} = \textbf{e}^{+}\,\frac{(1 + \text{cos}^{2}\iota)}{2}\frac{h_{\text{TT}}}{\omega}\text{cos}(2\omega t + 2\Phi_{0}) \nonumber \\
&& \textbf{H}^{\times} = \textbf{e}^{\times}\,\text{cos}\,\iota\frac{h_{\text{TT}}}{\omega}\text{sin}(2\omega t + 2\Phi_{0}) \nonumber \\
&& \textbf{H}^{\odot} = \textbf{e}^{\odot}\,\text{sin}\,\iota\frac{h_{\text{ST}}}{\omega}\text{cos}(\omega t + \Phi_{0}) \nonumber \\
&& \textbf{H}^{\text{u}} = \textbf{e}^{\text{u}}\,\text{cos}\,\iota\frac{h_{\text{VL}}}{\omega}\text{cos}(\omega t + \Phi_{0}) \nonumber \\
&& \textbf{H}^{\text{v}} = \textbf{e}^{\text{v}}\,\frac{h_{\text{VL}}}{\omega}\text{sin}(\omega t + \Phi_{0}) \nonumber \\
&& \textbf{H}^{\leftrightarrow} = \textbf{e}^{\leftrightarrow}\,\text{sin}\,\iota\frac{h_{\text{SL}}}{\omega}\text{cos}(\omega t + \Phi_{0}) ,
\end{eqnarray}
where
\begin{eqnarray}\label{altpolstrain}
&& h_{\text{TT}}=\frac{2 \mathcal{M}}{d_{L}}(\mathcal{M}\omega)^{2/3} \nonumber \\
&& h_{\text{ST}}=\alpha_{\text{ST}}\frac{ \mathcal{M}}{d_{L}}(\mathcal{M}\omega)^{1/3} \nonumber \\
&& h_{\text{VL}}=\alpha_{\text{VL}}\frac{ \mathcal{M}}{d_{L}}(\mathcal{M}\omega)^{1/3} \nonumber \\
&& h_{\text{SL}}=\alpha_{\text{SL}}\frac{ \mathcal{M}}{d_{L}}(\mathcal{M}\omega)^{1/3} 
\end{eqnarray}
and $\alpha_{\text{ST,VL,SL}}$ are dimensionless couplings coefficients, which we treat as independent. Our $h_{\text{TT}}$ is equivalent to $h_{0}$ in Eq.(20) of Ref.~\cite{NG11yrCW}. Note that we have neglected higher order post-Newtonian corrections to the amplitude. The gravitational coupling strength can be modified in alternative theories of gravity, but here we maintain the $G=1$ scaling and absorb any changes via the coupling coefficients $\alpha_{\text{ST,VL,SL}}$.

We work on the assumption that any alternative polarization states are dominated by dipole emission, which is to be expected unless special symmetries eliminate the dipole contribution. We have also assumed that the tensor and vector waves are elliptically polarized, which should be the case for the leading order emission from a circular binary. Note that we can define $h_{\text{TT}}$ to be positive, but we have to allow $h_{\text{ST}}$, $h_{\text{VL}}$, and $h_{\text{SL}}$ to range over negative and positive values as the dipole charges can be negative or positive. We can use the data to derive bounds on the absolute values of the amplitudes. 

In the GR case of Eq.(\ref{residual}), degeneracies exist in the timing residuals: $r^{GR}(\psi,\Phi_{0},\Phi_{p,0})=r^{GR}(\psi+\pi/2,\Phi_{0}+\pi/2,\Phi_{p,0}+\pi/2)$, and trivially $r^{GR}(\psi)=r^{GR}(\psi+\pi)$ and $r^{GR}(\Phi_{0},\Phi_{p,0})=r^{GR}(\Phi_{0}+\pi,\Phi_{p,0}+\pi)$. If the pulsar terms are considered unimportant noise, the transformation further simplifies. Similar degeneracies exist in our parameterization, namely $r(\psi,h_{VL})=r(\psi+\pi,-h_{VL})$, and $r(\Phi_{0},\Phi_{p,0},h_{ST},h_{SL},h_{VL})=r^{GR}(\Phi_{0}+\pi,\Phi_{p,0}+\pi,-h_{ST},-h_{SL},-h_{VL})$. The standard analysis in the GR case exploits these mappings to
restrict the prior ranges on the polarization and phase parameters. Here we are permitted to do the same, so long as we include the sign freedom in the strain amplitudes.

Ideally we would choose priors on the source parameters that are similar to those used in the standard GR analysis, but this is difficult to do since the dipole radiation introduces additional terms into the frequency evolution. To cover a large range of possibilities we adopt scale-invariant priors that are log uniform in $\alpha_{D}$ and $\alpha_{Q}$. In the GR case priors on the chirp mass $\mathcal{M}$ translate directly into priors on $\alpha_{Q}$. The additional polarization modes will contribute to the quadrupole emission so the mapping is modified in a theory-dependent fashion.
There is less guidance on what prior bounds to use for $\alpha_{D}$. One way to set boundaries on the prior range for these parameters is to impose self-consistency conditions. Our model assumes that the signals do not evolve significantly during the duration of the observation, which implies that $\dot{\omega}T_{\text{obs}}^{2} \ll 1$. The problem here is that even in the GR limit the self-consistency relation can be violated for the most massive systems at high frequencies. Setting $\dot{\omega}T_{\text{obs}}^{2}=1$ in the GR limit, we get
\begin{eqnarray}
&& \mathcal{M}_{\text{max}} = 4 \times 10^{7}M_{\odot} \Big( \frac{3\times 10^{-7}\text{Hz}}{f_{\text{max}}}   \Big)^{11/5} \Big( \frac{10\text{yr}}{T_{\text{obs}}} \Big)^{6/5}
\end{eqnarray}
Turning this around and using a minimum mass chirp mass of $\mathcal{M} = 10^{8}M_{\odot}$, the lowest value considered in the NANOGrav analyses, we see that the no-chirp condition is violated at $f=1.9\times10^{-7}$Hz for the lowest mass systems. For the highest mass systems with $\mathcal{M}=10^{10}M_{\odot}$, the no-chirp condition is violated at $f=2.4\times10^{-8}$Hz. Note that we are just requiring that the signal moves less than a frequency bin during the observations. The criteria really should be some small fraction of a bin. Imposing the bound at highest frequencies effectively limits the allowed chirp masses at all frequencies. The alternative is to change the model and allow for frequency evolution, at least during the pulsar-to-Earth pulse travel time. Then we need to integrate $\omega(t)$ with respect to time, which can be done by recasting the integrand to the following form:
\begin{eqnarray}\label{Phi(t)}
&& \Phi(t)-\Phi_{0} = \int_{t_{0}}^{t}\omega(t)dt = \int_{\omega_{0}}^{\omega(t)} \frac{d\omega}{\alpha_{D}\,\omega^{2}+\alpha_{Q}\,\omega^{8/3}} \nonumber \\
&& =3 \frac{\alpha_{Q}^{3/2}}{\alpha_{D}^{5/2}} \Bigg(\text{tan}^{-1}\Big( \sqrt{ \frac{\alpha_{Q}\, \omega(t)^{2/3}}{\alpha_{D}}} \Big) - \text{tan}^{-1}\Big(  \sqrt{\frac{\alpha_{Q}\, \omega_{0}^{2/3}}{\alpha_{D}}} \Big) \Bigg) \nonumber \\
&& + \frac{1}{\alpha_{D}} \Bigg(\frac{1}{\omega_{0}} - \frac{1}{\omega(t)} \Bigg) + \frac{3\,\alpha_{Q}}{\alpha_{D}^{2}} \Bigg( \frac{1}{\omega(t)^{1/3}} - \frac{1}{\omega_{0}^{1/3}}   \Bigg).
\end{eqnarray}
It is easier to understand the dipole correction in the PN framework when we take the limit $\alpha_D\rightarrow 0$ such that $\alpha_D \ll \alpha_Q \omega_0^{2/3}$, which to first order is:
\begin{eqnarray}
\Phi(t)-\Phi_{0} \approx \frac{3}{5\,\alpha_{Q}}\Big(\frac{1}{\omega_{0}^{5/3}}-\frac{1}{\omega(t)^{5/3}}\Big)- \nonumber \\
\frac{3\,\alpha_{D}}{7\,\alpha_{Q}^{2}}\Big(\frac{1}{\omega_{0}^{7/3}}-\frac{1}{\omega(t)^{7/3}}\Big)
\end{eqnarray}
We see how this corresponds to the GR case with a small correction. Unfortunately, the dipole term also makes $\omega(t)$ a transcendental function of time
\begin{eqnarray}\label{omega(t)}
&& t-t_{0} = \int_{\omega_{0}}^{\omega(t)} \frac{d\omega}{\alpha_{D}\,\omega^{3}+\alpha_{Q}\,\omega^{11/3}} \nonumber \\
&& = \frac{1}{4\,\alpha_{D}^{4}} \Bigg[ 2\,\alpha_{Q}^{3} \Bigg(2\,\text{ln}\Bigg( \frac{\omega_{0}}{\omega(t)} \Bigg) + 3\,\text{ln}\Bigg( \frac{\alpha_{D} + \alpha_{Q}\,\omega(t)^{2/3}}{\alpha_{D} + \alpha_{Q}\,\omega_{0}^{2/3}}  \Bigg) \nonumber \\
&& 2\,\alpha_{D}^{3} \Bigg(\frac{1}{\omega_{0}^{2}} - \frac{1}{\omega(t)^{2}} \Bigg)   +3\,\alpha_{Q}\,\alpha_{D}^{2} \Bigg( \frac{1}{\omega(t)^{4/3}} - \frac{1}{\omega_{0}^{4/3}}     \Bigg) \nonumber \\
&& \hspace{1.0 in} + 6\,\alpha_{Q}^{2}\,\alpha_{D} \Bigg(  \frac{1}{\omega_{0}^{2/3}} - \frac{1}{\omega(t)^{2/3}}   \Bigg)   \Bigg) \Bigg]
\end{eqnarray}
With the full evolution included, it is less clear how we should choose maximum values for $\alpha_{D}$, $\alpha_{Q}$. One extreme might be to demand that the systems do not merge during the observation time, which we can define as when the Earth term frequency becomes infinite during the observation time:
\begin{eqnarray}
&& T_{\text{merge}} = \frac{3\,\alpha_{Q}^{3}}{2\,\alpha_{D}^{4}}\Big[\text{ln}\Big(\frac{\alpha_{Q}\,\omega_{0}^{2/3}}{\alpha_{D} + \alpha_{Q}\,\omega_{0}^{2/3}}\Big) \Big] + \frac{1}{2\,\alpha_{D}\,\omega_{0}^{2}}
\nonumber \\
&& \hspace{1.0 in}  -\frac{3\,\alpha_{Q}}{4\,\alpha_{D}^{2}\,\omega_{0}^{4/3}} + \frac{3\,\alpha_{Q}^{2}}{2\,\alpha_{D}^{3}\,\omega^{2/3}}
\end{eqnarray}
In the context of GR, this corresponds to the condition $T_{\text{merge}} < T_{\text{obs}}$ where
\begin{eqnarray}
&& T_{\text{merge}}^{GR} = \frac{5}{256}\mathcal{M}^{-5/3}\omega^{-8/3} \nonumber \\
&& = 2\, \text{years} \Big( \frac{10^{10}M_{\odot}}{\mathcal{M}}   \Big)^{-5/3}  \Big( \frac{10^{-7}\text{Hz}}{f}   \Big)^{-8/3}
\end{eqnarray}
More generally we can define $T_{\text{chirp}} = \omega/\dot{\omega}$. This quantity is similar to $T_{\text{merge}}$ but is easier to compute for modified theories (in GR, $T_{\text{chirp}} = \frac{8}{3} T_{\text{merge}}$). Treating the dipole and quadrupole extremes separately we have the limits
\begin{eqnarray}\label{EvolveUpperBound}
&& \alpha_{D} < \frac{1}{\omega_{0}^{2}\,T_{\text{obs}}} \nonumber \\
&& \alpha_{Q} < \frac{1}{\omega_{0}^{8/3}\,T_{\text{obs}}}
\end{eqnarray}
Here $\omega_{0}$ is the initial orbital angular frequency at the Earth. The merger time is related to the chirp time by a factor less than unity, so
we multiply Eq.(\ref{EvolveUpperBound}) by a factor of one tenth to define the no-merger condition, which then defines the upper bounds on $\alpha_D$ and $\alpha_Q$. For the lower limits we can choose values that we know apriori produce un-observable frequency changes: $\dot{\omega}L T_{\text{obs}} \ll 1$. Treating the dipole and quadrupole extremes separately, we have the limits
\begin{eqnarray}
&& \alpha_{D} > \frac{1}{\omega_{0}^{3}\,T_{\text{obs}}\,L} \nonumber \\
&& \alpha_{Q} > \frac{1}{\omega_{0}^{11/3}\,T_{\text{obs}}\,L}
\end{eqnarray}
In other words, the GW becomes effectively monochromatic, with the pulsar frequency roughly equal to the Earth term frequency. We will need to use priors that depend on the Earth term frequency. Note that maximum values are a factor of $\omega_{0}L$ larger than the minimum values. Consequently the prior range on the frequency evolution will be very different from the GR case, and this will impact the the upper limits. Note that depending on the choice of the radiative-loss coupling and the GW frequency, it is possible that just the pulsar term or just the Earth term falls in the observation band.

Since the distance to each pulsar $L_i$ is not known to high precision, we marginalize over the distance to each pulsar. In principle the orbital phase seen at each pulsar, $\Phi_{i}$, is determined by the time delay $L_{i}(1 + \bm{\Omega}\cdot\textbf{p})$, but since the $L_{i}$ are not well constrained we get phase wrapping in the pulsar signals that makes it very difficult to marginalize over the $L_i$. One way around this is to introduce independent phase
terms $\Phi_{i}$ for each pulsar~\cite{2010arXiv1008.1782C}, and only keeping the $L_{i}$ dependence in the pulsar frequencies. In effect this splits the pulsar distance into two parts, a large part on the order of 1 kpc, and a small correction of order $2\pi/\omega \sim 1$ pc.

\section{Data Analysis Methods}\label{section3}
To simulate the pulsar timing data, we used the \texttt{libstempo}\footnote{https://github.com/vallis/libstempo} package \texttt{toasim.py} to generate the timing residuals for each pulsar. Using the 11 year data release pulsars~\cite{2018ApJS..235...37A} with ephemeris DE435 and the \texttt{fake\_pulsar} function, we created a mock pulsar data set with an 11 year observation time at a 30 day cadence with random 1 day offsets to mimic the irregularity of pulsar observations. For all realizations, the TOA uncertainty assigned to all pulsars was $\sigma_{\text{TOA}}=0.5\mu$s. We then added white noise of EFAC $=1$ and $\sigma_{\text{EQUAD}}=100$ns, where the total rms white noise is defined as $\sigma^{2} = (\text{EFAC})^{2}\sigma^{2}_{\text{TOA}} + \sigma^{2}_{\text{EQUAD}}$. To the pulsar distances we added a random Gaussian variate, proportional to its cataloged uncertainty. For pulsars whose distance is not well known, we used a distance of 1 kpc and assigned a 20$\%$ error in the distance. We then added a random uniform component proportional to the GW wavelength relative to the pulsar distance in order properly de-phase the pulsar and Earth terms. 

For simulations with a GW signal, we used a modified version of the function \texttt{create\_cw} to generate a continuous wave signal based on the model outlined in Section~\ref{section2}. We used a fixed quadrupolar GW frequency $f^{\text{TT}}_{\text{GW}} = 1\times10^{-8}$Hz to ensure both the quadrupolar and dipolar signatures appear in the observation band at roughly the same sensitivity. This corresponds to an initial Earth term orbital frequency $\omega_{0} = \pi f^{\text{TT}}_{\text{GW}}$. For all simulations we chose $\psi=\pi/4$, $\Phi_{0}=\pi/4$, and $\text{cos}\,\iota=0.5$.

Ignoring the timing model for now and considering only white noise, the total SNR of an alt-pol signal injection is equal to\cite{Rosado:2015epa}:
\begin{eqnarray}\label{SNR}
    \text{SNR}^{2}_{\rm inj} = \frac{2}{S_{n}}\sum\limits^{N_{\text{puls}}}_{i} \int_{0}^{T_{\text{obs}}}dt\,(\sum\limits_{A} r^{A}_{i}(t_{e}))^{2} \nonumber \\
    \equiv\frac{1}{\sigma^{2}}\sum\limits^{N_{\text{puls}}}_{i} \sum\limits_{n}^{N_{\text{TOA}}}(\sum\limits_{A} r^{A}_{i}(t_{n}))^{2}
\end{eqnarray}
where $ r^{A}_{i}(t_{e})$ is defined by Eq.(\ref{residual}), $t_{n}$ is the $n$th TOA, and $S_{n}=2\sigma^{2} \Delta t$, where $\Delta t$ is the cadence. The sums are over the polarizations and pulsars. We normalize our injections using this definition, with each mode contributing roughly equal SNR$^{2}$; however, it is difficult to gauge what constitutes equal considering the complicated form of Eq.(\ref{SNR}) when substituting in Eq.(\ref{residual}). We simply choose a target SNR and find the values of the amplitudes, individually, that achieve this target for a given sky location. We then add all the amplitudes together and rescale them to achieve the target SNR. Note that the normalization will be dependent on the configuration of the array, particularly with respect to longitudinal modes. If a SL longitudinal signal were directly behind a particular pulsar, virtually all SNR information is contained in that pulsar's residuals per our normalization procedure. If there were no enhancement but there were still many pulsars near the GW source, then the longitudinal amplitudes would be determined by the collective SNR of those pulsars, on the same order as transverse modes. If we had pulsars only in the sky region opposite to the GW source, the response would be reduced, requiring us to inject an appreciably louder signal. 

We should emphasize that while we have normalized the injections according to Eq.(\ref{SNR}), this definition is valid only for higher frequencies in the band because the fitting of the timing model in Eq.(\ref{sigmodel}) reduces the sensitivity at lower frequencies, and this effects the SNR. We found the effective SNR by empirically computing the likelihood ratio by dividing the maximum likelihood value by the likelihood when the GW amplitudes are set equal to zero. The log-likelihood ratio $\Lambda$ scales with the measured SNR as $\Lambda=\text{SNR}_{\rm eff}^{2}/2$, and we use this relation to define the effective SNR of the signal. See Table~\ref{tab:CW} for the corresponding effective and injected SNRs.

We used the same likelihood and Bayesian framework outlined in Ref.~\cite{0004-637X-794-2-141}, which used NANOGrav's software package \texttt{enterprise}\footnote{https://github.com/nanograv/enterprise} to implement the search with \texttt{PTMCMCSampler}\footnote{https://github.com/jellis18/PTMCMCSampler}. The common parameters in our search are indicated in Eq.(\ref{altpolparam}). All non-amplitude parameter priors were uniform, except for the evolution couplings which were log uniform. For the alt-pol signals we added parameters that allowed the sign of the alt-pol strain amplitudes to be positive or negative. We assumed Gaussian priors on the pulsar distances $L_i$, centered on the observed value and with a standard deviation given by the measured uncertainty. A uniform prior was assumed for the pulsar initial phase terms.

We discuss seven main analyses, the first six of which are outlined in Table~\ref{tab:CW}. The first analysis is on a simulated data set with all polarizations present, and uses the full polarization model to recover the signal. The purpose of this analysis is to test the analysis software and see how well the various model parameters can be recovered. The second analysis uses simulated data where a pure TT signal has been injected, and the recovery is done using each polarization individually. The third analysis is on pure noise realizations with the goal of comparing the upper limits that can be placed on the amplitude of each polarization mode. The fourth analysis uses the same noise-only data, but with the sky location of the source restricted to be near pulsar J1024-0719 at (-0.13, 2.72); the choice was incidental as this pulsar was closest to one of our signal injections. The goal here is to investigate how the enhanced response in the longitudinal modes tends to push the inferred sky localization away from the pulsar locations. The fifth and sixth analyses use simulated data with a pure TT signal, and look at the upper limits that can be placed on the alt-pol modes when a TT signal is detected. The data sets differ in the sky location of the source, with the source for analysis six
placed in the direction of the galactic center (GC), where the array has more pulsars. The seventh and final analysis uses noise-only data with noise levels similar to those found in contemporary timing arrays to produce upper limits as a function of sky location.

\begin{table*}[htp]
	\begin{tabular}{
	 |p{3.5cm}|p{0.9cm}||p{2.0cm}|p{1.1cm}|p{1.2cm}|p{1.2cm}|p{2cm}|p{2.4cm}|p{1.0cm}|}
		\hline
		\multicolumn{9}{|c|}{Data Simulation Parameters} \\
		\hline
		Analysis Model& Index &Strain Priors & Injected Strains & Injected SNR & Effective SNR  & Sky Location ($\text{cos}\,\theta,\phi$) & Injected Radiative-Loss & Figures \\
		\hline
		\hline
		All modes& 1 & log-uniform   & TT, ST, SL, VL & 40 & $\sim$20   &(-.468, 4.647) &  $\text{log}_{10}(\alpha_{D})=3.5$ $\text{log}_{10}(\alpha_{Q})=8.5$ &  \ref{fig:GCSNR40strain}, \ref{fig:GCSNR40sky}, \ref{fig:GCSNR40evol} \\
		\hline
		Single mode only& 2 & log-uniform   & TT & 20 & $\sim$10   &(0, $\pi$) & $\text{log}_{10}(\alpha_{Q})=8.5$ & \ref{fig:STsearch}  \\
		\hline
		All modes, sky-averaged &3 & uniform  & - & 0 &0 &  - & - & -  \\
		\hline
		All modes, restricted around J1024-0719&4 & uniform  & - & 0 &0   &  (-.13,2.72 ) & - & \ref{fig:sky_UL_NoSignalBehind} \\
		\hline
		All modes&5 & uniform & TT & 20 & $\sim$10   & (0, $\pi$) & $\text{log}_{10}(\alpha_{Q})=8.5$ & \ref{fig:strain_UL_SNR20TT} \\
		\hline
		All modes&6 & uniform & TT & 20 & $\sim$10   & (-.468, 4.647) & $\text{log}_{10}(\alpha_{Q})=8.5$ & - \\
		\hline
	\end{tabular}
	\caption{The set-up for each data simulation and analysis}
	\label{tab:CW}
\end{table*}

\section{Results}\label{sectionresults}

\begin{figure}[htp]
	{\includegraphics[width = 3.25in]{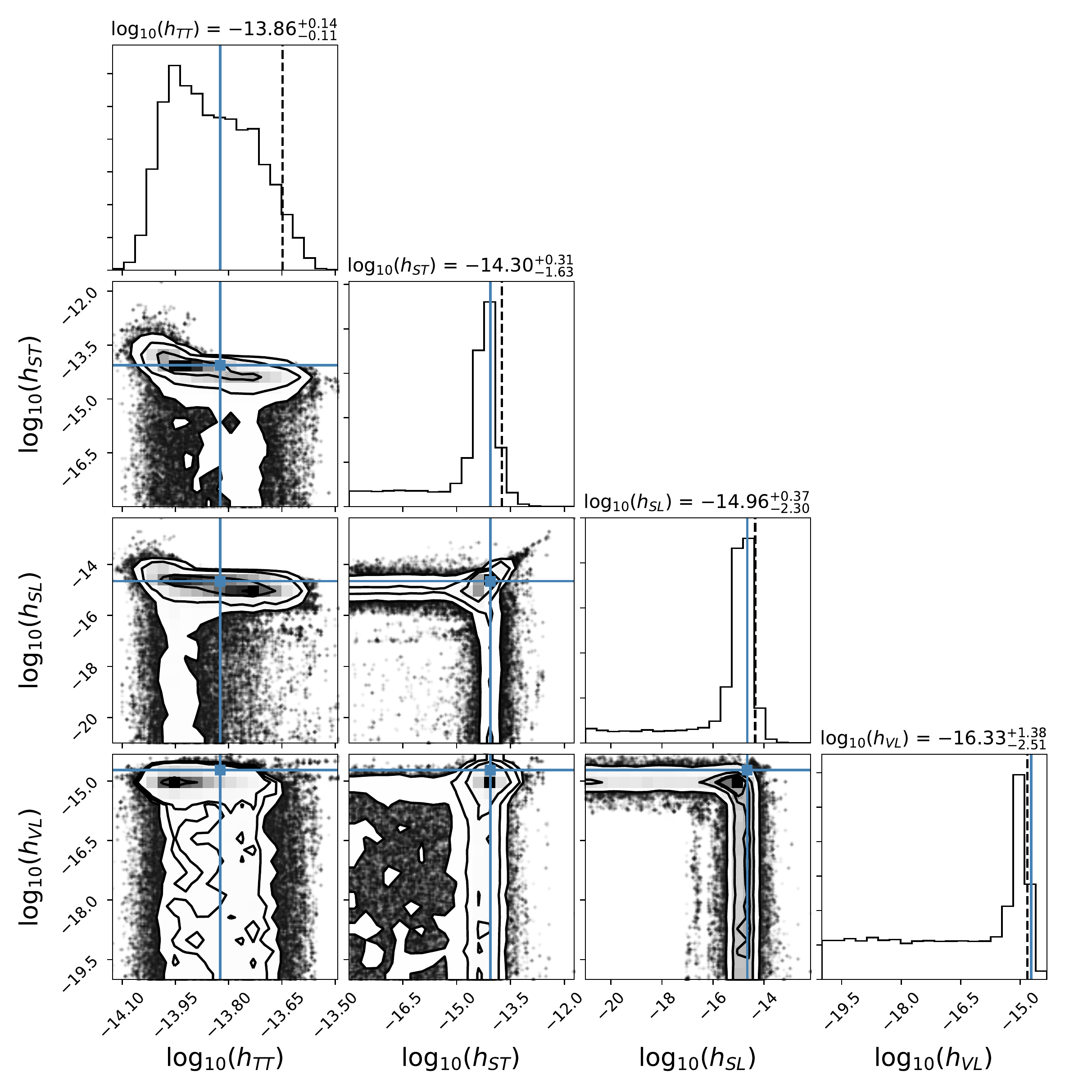}} 
	\caption{Joint posteriors for the strain parameters of an all modes search. The injected signal has all modes present and a total effective SNR$\sim20$. The priors for the strains are log-uniform. The blue lines are the injected values of the strains, and the dotted lines are 95$\%$ quantile. }
	\label{fig:GCSNR40strain}
\end{figure}

\begin{figure}[htp]
	{\includegraphics[width = 3.25in]{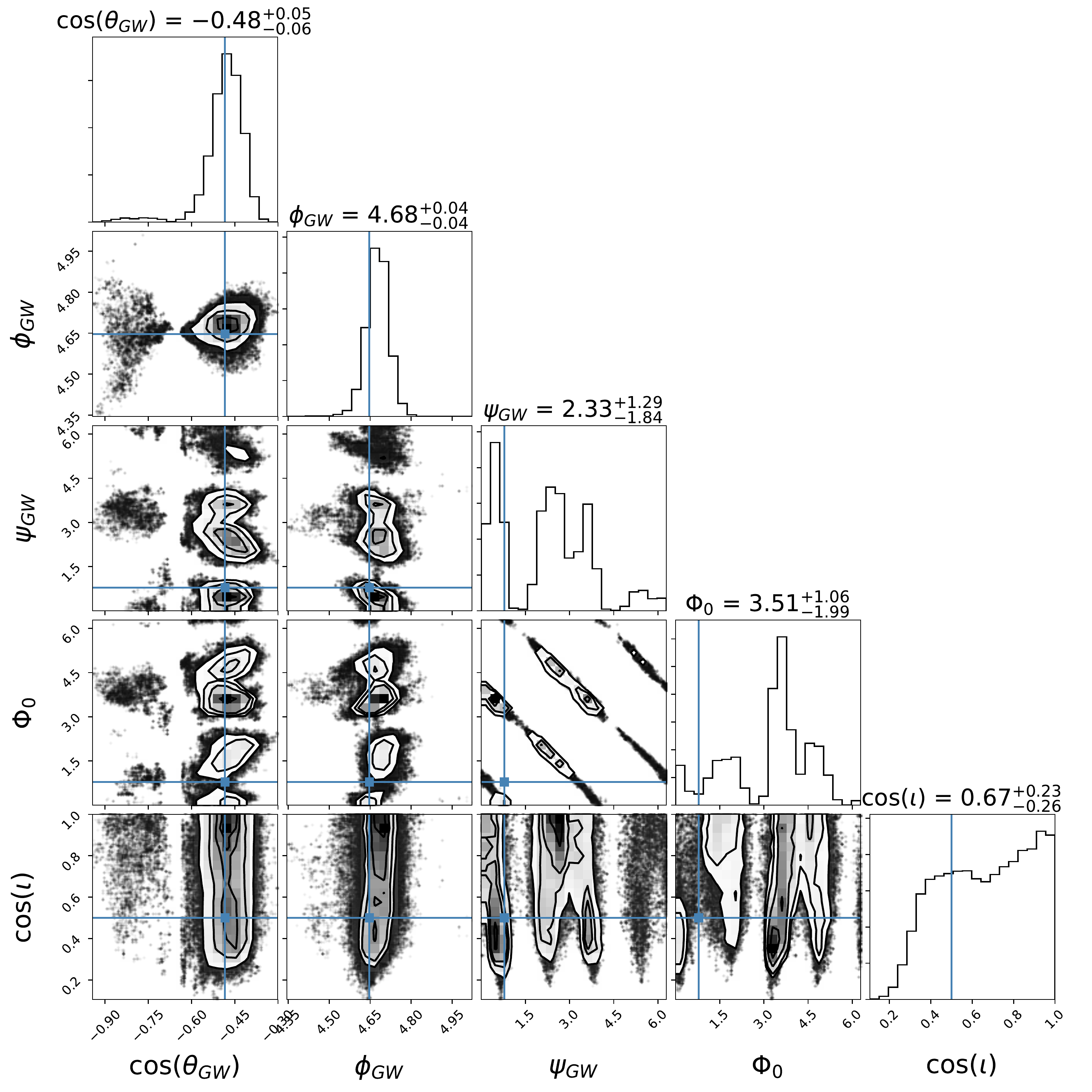}} 
	\caption{Joint posteriors for the sky parameters of an all modes search. The injected signal has all modes present and a total effective SNR$\sim20$. The priors for the parameters shown here are all uniform. The blue lines are the injected values of the parameters. The GW is originating from behind the GC. }
	\label{fig:GCSNR40sky}
\end{figure}

\begin{figure}[htp]
	{\includegraphics[width = 3.25in]{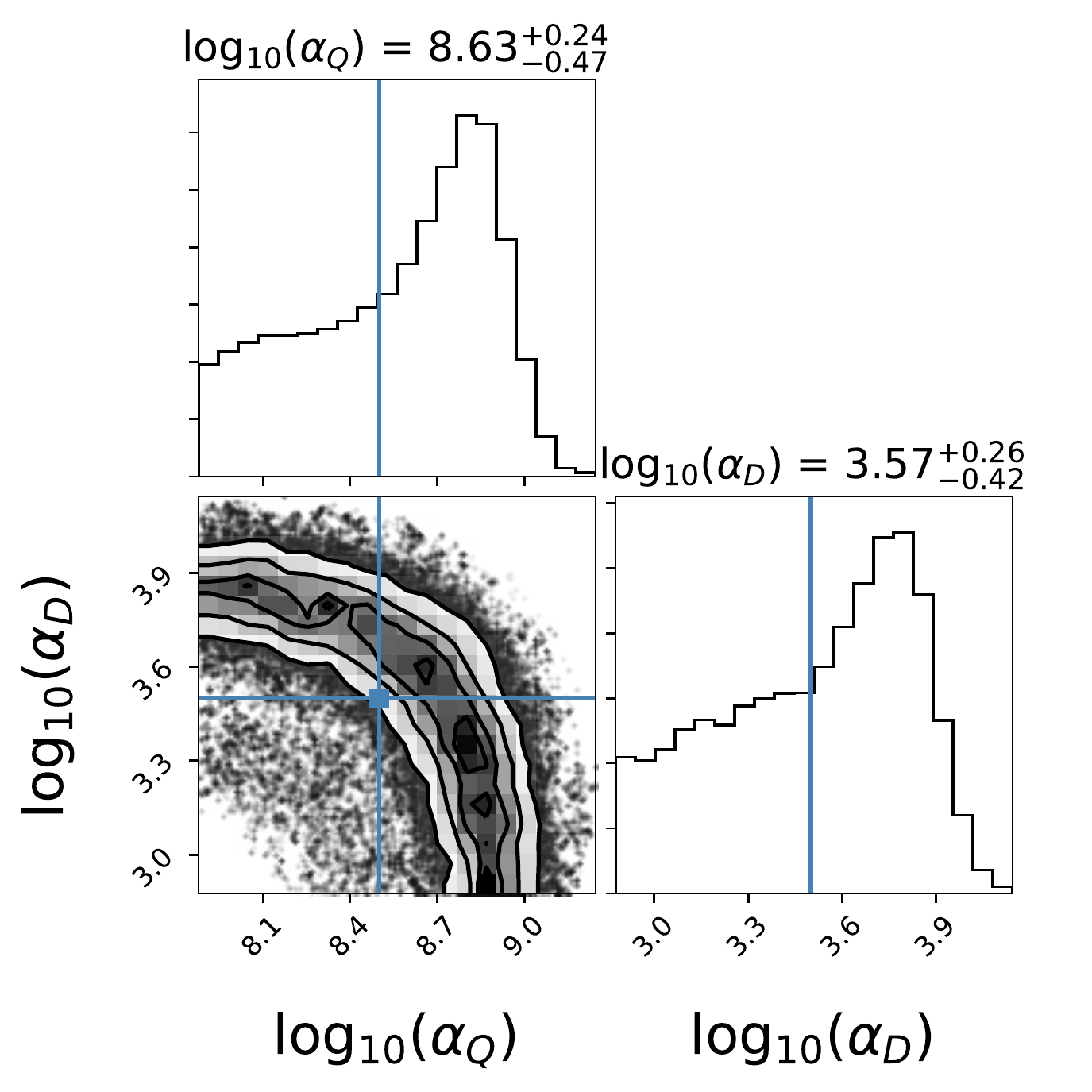}} 
	\caption{Joint posterior for the radiative-loss coupling parameters of an all modes search. The injected signal has all modes present and a total effective SNR$\sim20$. The priors for the parameters shown here are log-uniform. The blue lines are the injected values of the parameters. }
	\label{fig:GCSNR40evol}
\end{figure}

The results of each analysis are described in the following subsections. The figure summary is as follows:  Figures~\ref{fig:GCSNR40strain},~\ref{fig:GCSNR40sky}, and~\ref{fig:GCSNR40evol} show the alt-pol parameter recovery for simulated signals with effective signal-to-noise ratio $\text{SNR}_{\text{eff}}\sim20$. %Figure~\ref{fig:J1747-4036_p_phase} focuses on the recovery of the pulsar phase term for a pulsar with a similar sky location to the GW source. 
Figure~\ref{fig:STsearch} shows the results of a ST-only model search for simulated data with a TT-only signal. The ST model is able to recover the TT signal, but with biases in some parameters.
Figure~\ref{fig:sky_UL_NoSignalBehind} illustrates how the pulsar locations impact the upper limits analysis by creating ``zones of avoidance'' around the pulsar locations when no signal is present in the data.
Figure~\ref{fig:strain_UL_SNR20TT} explores the limits that could be placed on the alt-pol amplitudes in likely event that the signal is a pure GR TT-mode. Figure~\ref{fig:fixedsky} shows the amplitude upper limits as a function of sky location for each of the polarization modes when no signal is in the data.


\subsection*{Analysis 1: Full Alt-Pol Parameter Recovery}\label{sub1} 
The full parameter recovery of alt-pol searches validates our ability to probe higher dimensional signals. An interesting and unexpected covariance is seen between the SL and ST modes as well as the SL and VL modes in Figure~\ref{fig:GCSNR40strain}; apparently there exists a geometric degeneracy with respect to the array. This is an important aspect to account for in future studies. We have verified this disappears if we have a large enough amplitudes to discriminate the angle of inclination $\iota$. Note that the dipole radiation is far less sensitive for these frequencies than we normalized in the injection. Sky localization is very reliable, and the mapped posterior for other parameters is reasonable although we can see that the posterior has local maxima away from the true injected values with some unexpected structure. Fitting for the timing model can have an effect on some of these other parameter posteriors, such as $\Phi_{0}$ and $\text{cos}\,\iota$ along with the amplitudes because the fitting procedure changes the shape of the waveform, as seen visually in Ref.\cite{2012ApJ...756..175E}. Since the dipole signal exists at 5nHz, the lower order harmonic and any couplings to it are more effected by this than the quadrupole signal. Again, the effect becomes negligible for larger amplitudes. All evolution coupling posteriors appear like Figure~\ref{fig:GCSNR40evol}, regardless of the nature of the injection. The reason for this is that while a mixed dipole/quadrupole injection yields the transcendental function of Eq.(\ref{omega(t)}), the uncertainty in the pulsar distances allows a level of degeneracy with the non-mixed injections, and vice versa.

It should be noted that alt-pol injections render posterior modes in the pulsar phase parameters of pulsars near the GW source; all other pulsar phases sample uniformly. This is because the nearby pulsars dominate the SNR contribution of the longitudinal modes, so the pulsar phase becomes necessary to accurately describe the antenna pattern. 
%See Figure~\ref{fig:J1747-4036_p_phase}, which shows the pulsar phase of J1747-4036, the closest pulsar to an alt-pol GW originating behind the GC.
%\begin{figure}[htp]
%	{\includegraphics[width = 3.25in]{Plots/J1747-4036_cgw_p_phase_hist_GCSNR40Det.png}} 
%	\caption{Posterior of $\Phi_{J1747-4036}$ (in radians) for an all modes search. The injected signal, which originates very near this pulsar, has all modes present and a total effective SNR$\sim20$. The prior for the pulsar phase shown here is uniform. Notice that this pulsar phase posterior has 2 degenerate modes because this term becomes important for the response of longitudinal polarizations. Without longitudinal polarizations, this parameter would sample uniformly. }
%	\label{fig:J1747-4036_p_phase}
%\end{figure}

For very loud signals (SNR$_{\text{eff}}>$100), the signs of the dipole charges lock on to the true values. However, for moderate to low SNR, the posterior is not very sensitive to the value of the sign, and frequently accepts jumps to opposite signed values because the difference in the likelihood is not very significant. As a check, we restricted the values of the signs to the true injected values and did not find an appreciable difference between the posterior distributions compared to the marginalized search. However, the periodic covariance between $\Phi_{0}$ and $\psi$ in Figure~\ref{fig:GCSNR40sky} is a result of the marginalization of the signs of the dipole strain and is much more selective in the restricted case.

\subsection*{Analysis 2: Single Polarization Recovery}\label{sub2}

\begin{figure}[htp]
	{\includegraphics[width = 3.25in]{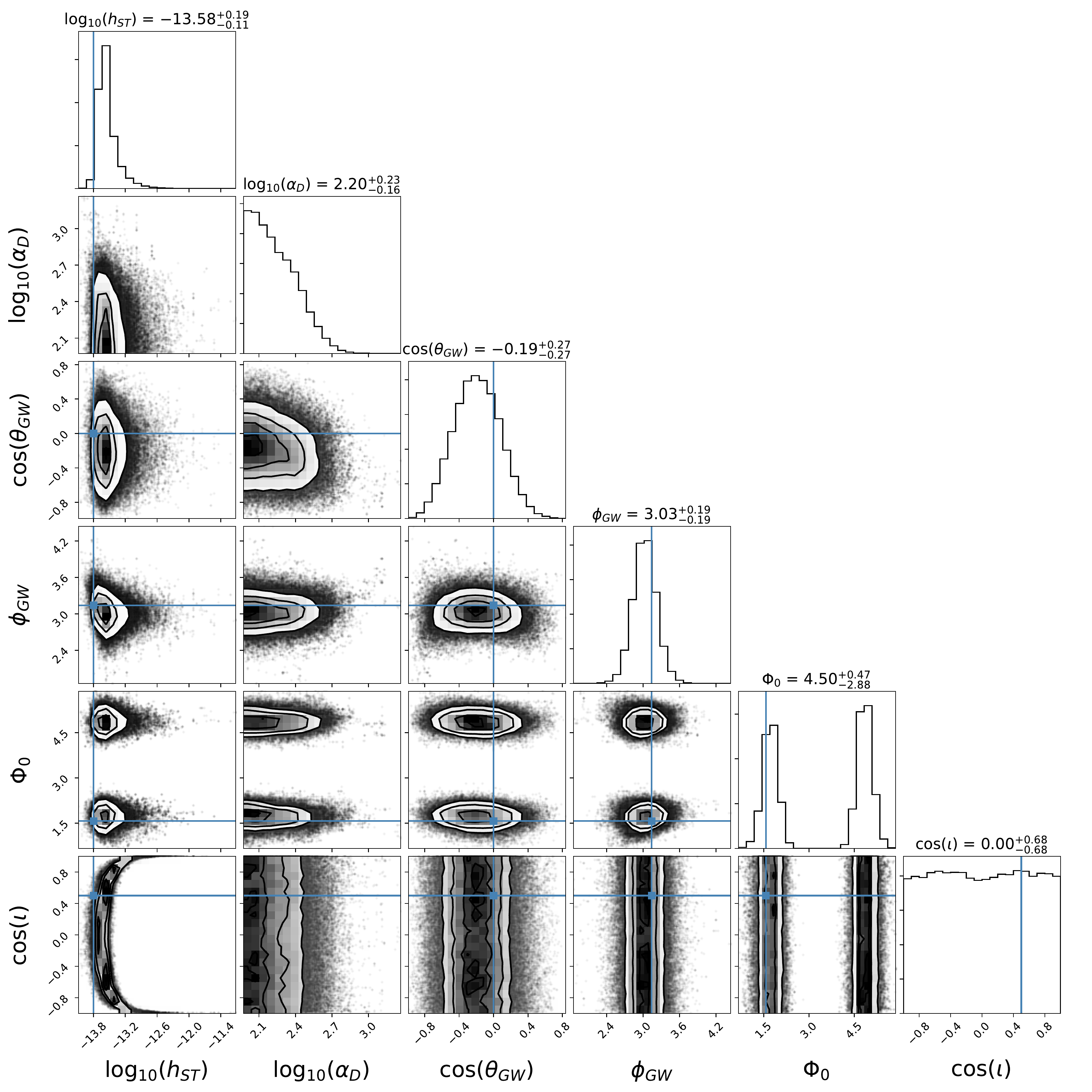}} 
	\caption{Joint posteriors for an analysis using a pure ST model on data with a pure TT signal with SNR$_\text{eff}\sim10$ at $f_{\text{GW}}=1\times10^{-8}$Hz. The ST model is able to recover the TT signal, albeit with a biased $\alpha_{D}$ and amplitude. Identical analyses with longitudinal modes render biases in the sky localization. Two truth lines have been modified. Eq.(\ref{altpolresiduals}) shows that $\Phi_{0}=\pi/4$ for a TT mode injection will be seen as $\Phi_{0}=\pi/2$ for an ST mode that is in phase with that injection, and we have reflected that here. Also, since $\log_{10}{\alpha_{Q}}=8.5$ for this injection, we recast the truth value in terms of the correct units, $\log_{10}{\alpha_{Q}\omega^{2/3}}=\log_{10}{\alpha_{D}}=3.5$, which is outside the posterior. }
	\label{fig:STsearch}
\end{figure}

To test if one signal could be mistaken for another, we searched a ST GW in phase with an injected TT mode of SNR$_\text{eff}\sim10$ at $f_{\text{GW}}=1\times10^{-8}$Hz. The physical motivation for this lies in the possibility of superluminal alt-pol GWs, which could arrive far earlier than the TT part of the signal, prompting an individual polarization search similar to tests performed on LIGO/Virgo data~\cite{Abbott:2017tlp,Abbott:2017oio,Abbott:2018utx}. Figure~\ref{fig:STsearch} shows the results of one such analysis. We see that a pure ST model can recover a pure TT signal, albeit with a biased amplitude and $\alpha_{D}$. 
We found that any single polarization search can yield a detection of a TT signal (with a biased sky localization for longitudinal modes), but that when all modes are included in the model the correct TT model is preferred (see Analysis 5 below). The longitudinal modes recover less of the TT signal and have poorer sky localization than either of the transverse modes because the response functions are very different. Incidentally, the ST search sky location and amplitudes are close to the injected TT mode values due to the array having similar geometric sensitivity to any transverse mode. We find the mapped posteriors agree well with the injected parameters of the TT signal when searching for the TT mode only.

\subsection*{Analysis 3: Sky-Averaged Upper Limits}\label{sub3}
For the following Analyses 3 and 4 we simulated noise-only data to perform an upper limit search. If we conjecture that a signal is present yet undetectable, we want to know the largest values the amplitudes can be. To this end we use uniform amplitude priors for upper limits rather than log-uniform. The other parameter priors are still uniform. We performed a marginalized search of the upper limits of all polarizations. While it is known that even for the TT mode a sky-averaged upper limit yields a sky location bias, the inclusion of longitudinal modes greatly exaggerates this effect. In the absence of any signal, the posterior is diminished in sky regions with many pulsars because the enhanced sensitivity to longitudinal modes forces the amplitudes to lower values. Since the posterior is the product of the likelihood and the prior, the likelihood will be no different at lower amplitudes, but the prior will penalize them, preferring the largest values possible, thus rendering the bias in sky location. For the simulated NANOGrav array with $0.5\mu{\rm s}$ timing residuals we find sky-averaged upper limits of $h^{95\%}_{\text{TT}}<2.0\times10^{-14}, h^{95\%}_{\text{ST}}<1.3\times10^{-14}, h^{95\%}_{\text{VL}}<8.6\times10^{-15}$, and $h^{95\%}_{\text{SL}}<4.0\times10^{-14}$. Note that the projected limit on the TT mode is comparable to that found in the NANOGrav 11 year analysis~\cite{NG11yrCW}, and we anticipate that the same data set will yield bounds on the other modes that are in line with these simulated upper limits.

The sky location posteriors exhibit maxima at values where the response can remain geometrically hidden from the detector. The sky-averaged case provides a general proxy for the array's sensitivity to localized GWs.
\begin{figure}[htp]
	{\includegraphics[width = 3.25in]{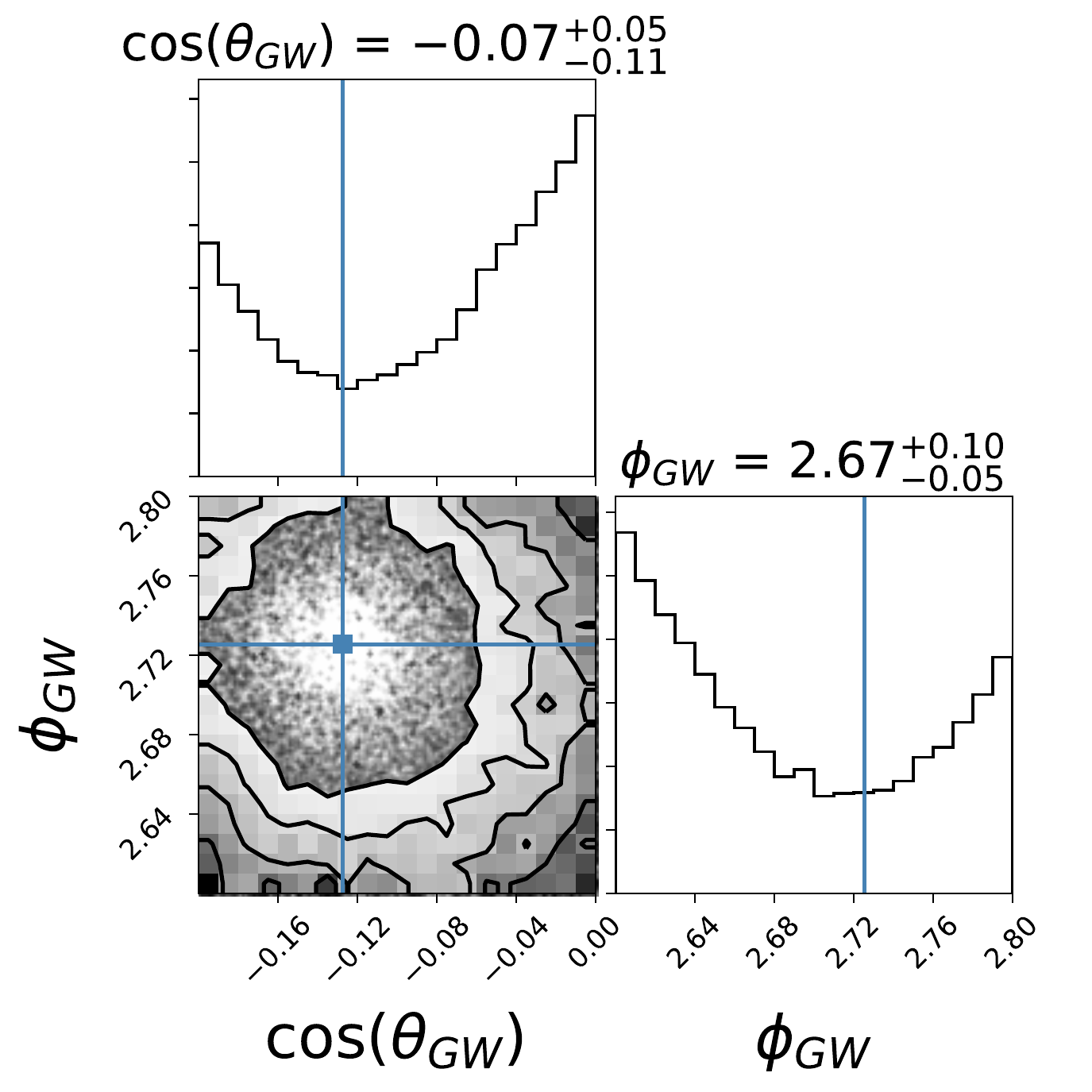}} 
	\caption{Joint posteriors of the sky location for an all modes upper limit search. No signal is present in this injection. All parameter priors are uniform except for $\alpha_{D}$ and $\alpha_{Q}$, which are log-uniform. The sky location is restricted in a region around J1024-0719, whose position is indicated in blue lines. The enhanced response from longitudinal modes causes the posterior to peak away from the pulsar.}
	\label{fig:sky_UL_NoSignalBehind}
\end{figure}
\subsection*{Analysis 4: Sky-Restricted Upper Limits}\label{sub4}
To further understand the nature of sky location bias, we performed a search nearly identical to Analysis 3, but restricted the sky location close to a pulsar, incidentally J1024-0719 in this case. Figure~\ref{fig:sky_UL_NoSignalBehind} shows the resulting sky location posterior, which is peaked away from the pulsar, and we found the resulting upper limit on the SL mode dramatically reduced compared to the sky-averaged case. The VL mode is not as severely effected as the pulsar is close to the less sensitive region directly aligned with the GW source, seen in Figure~\ref{fig:antenna}. Again, the other parameter posteriors exhibit maxima where the response can remain hidden from the detector. This analysis confirmed that the enhanced response from pulsars was dictating the shape of the upper limit posteriors.

\subsection*{Analyses 5 and 6: Alt-Pol Upper Limits with TT Injection }\label{sub56}

\begin{figure}[htp]
	{\includegraphics[width = 3.25in]{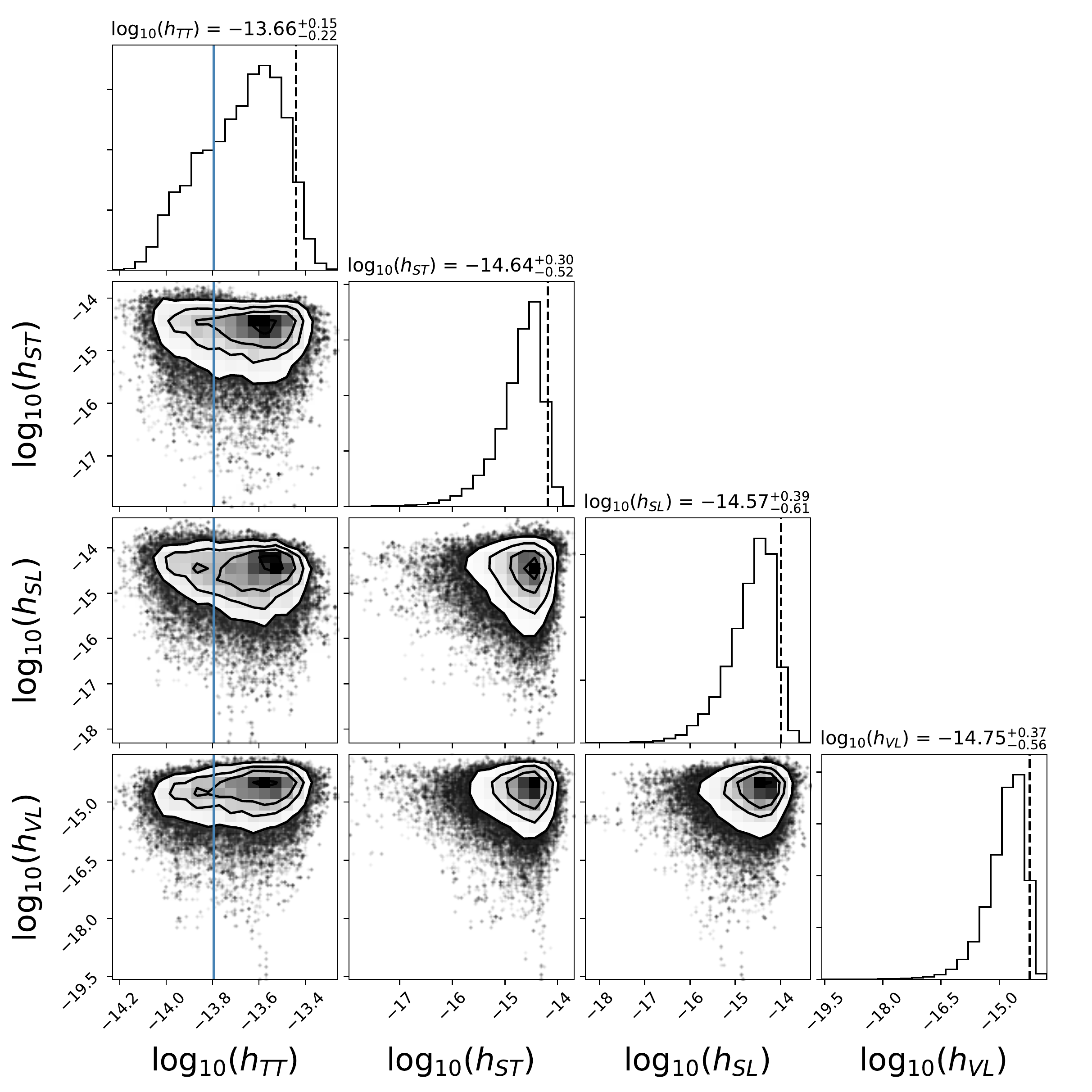}} 
	\caption{Joint posteriors of the strain upper limit strains in the presence of a pure GR signal with SNR$_\text{eff}\sim10$ located at (0,$\pi$).}
	\label{fig:strain_UL_SNR20TT}
\end{figure}

We also performed alt-pol upper limit searches in the presence of a TT mode injection with SNR$_\text{eff}\sim10$ for two separate sky locations, indicated in Table~\ref{tab:CW}. The idea here is that the detection of the TT-mode would constrain the orbital frequency and sky location of the source, potentially resulting in stronger bounds on the alt-pol modes. Unfortunately this was not found to be the case. The joint posteriors for the amplitude parameters for Analysis 5 are shown in Figure~\ref{fig:strain_UL_SNR20TT}. We found the resulting TT parameter posteriors recover the injected parameters as they did in Analysis 2, and that the alt-pol upper limits depend only on the sky location of the GW; the upper limits are more constrained if many pulsars are localized near the GW source. We find only for TT signals with SNR$_\text{eff}$ of order unity do we get the aforementioned sky location bias mentioned in Analyses 3 and 4; the likelihood's preference for the correct sky location is less significant in this case and starts to lose out to the alt-pol prior's avoidance of nearby pulsars. 

\begin{figure*}[t]
	{\includegraphics[width = 3.25in]{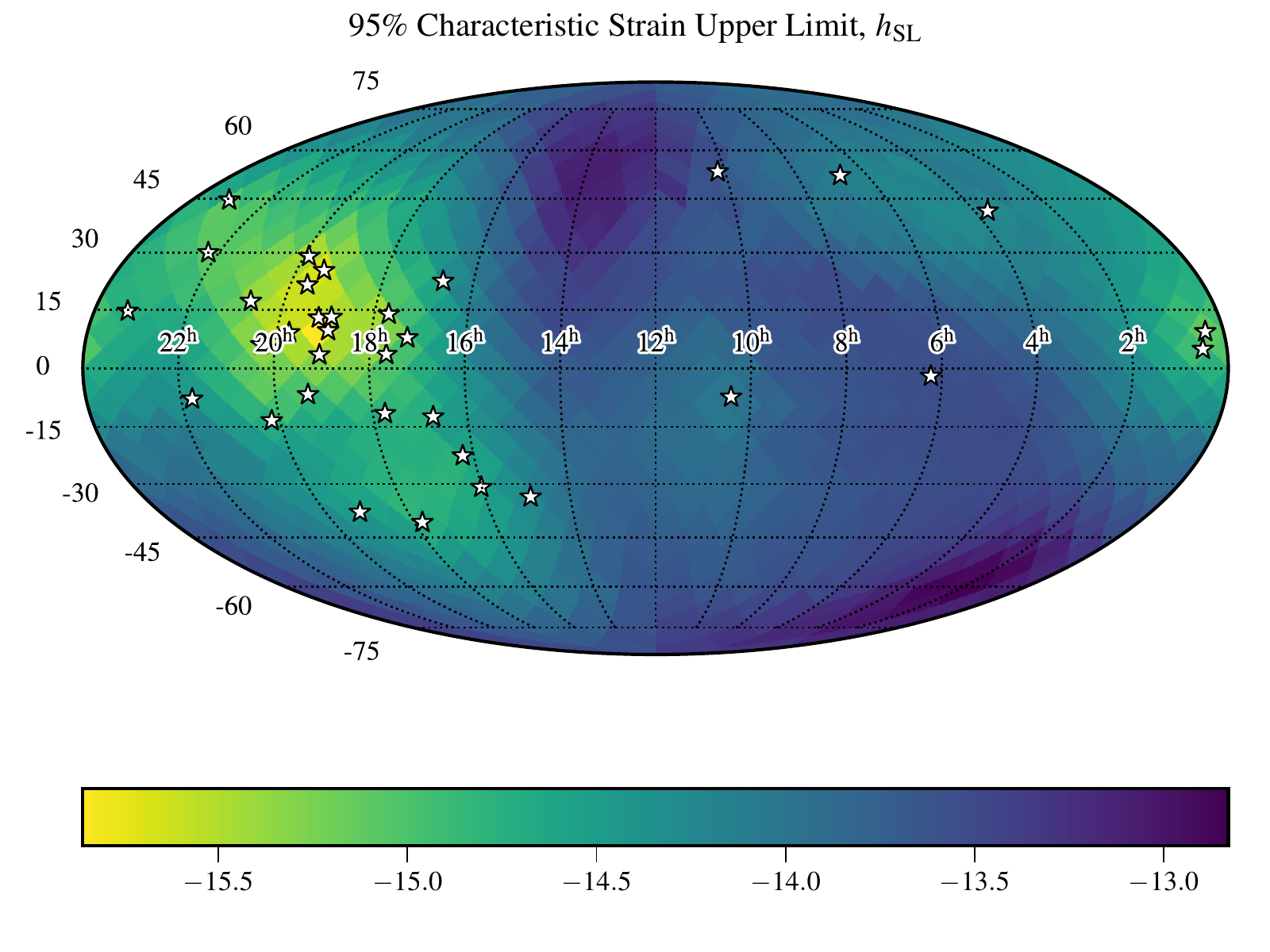}}  
	{\includegraphics[width = 3.25in]{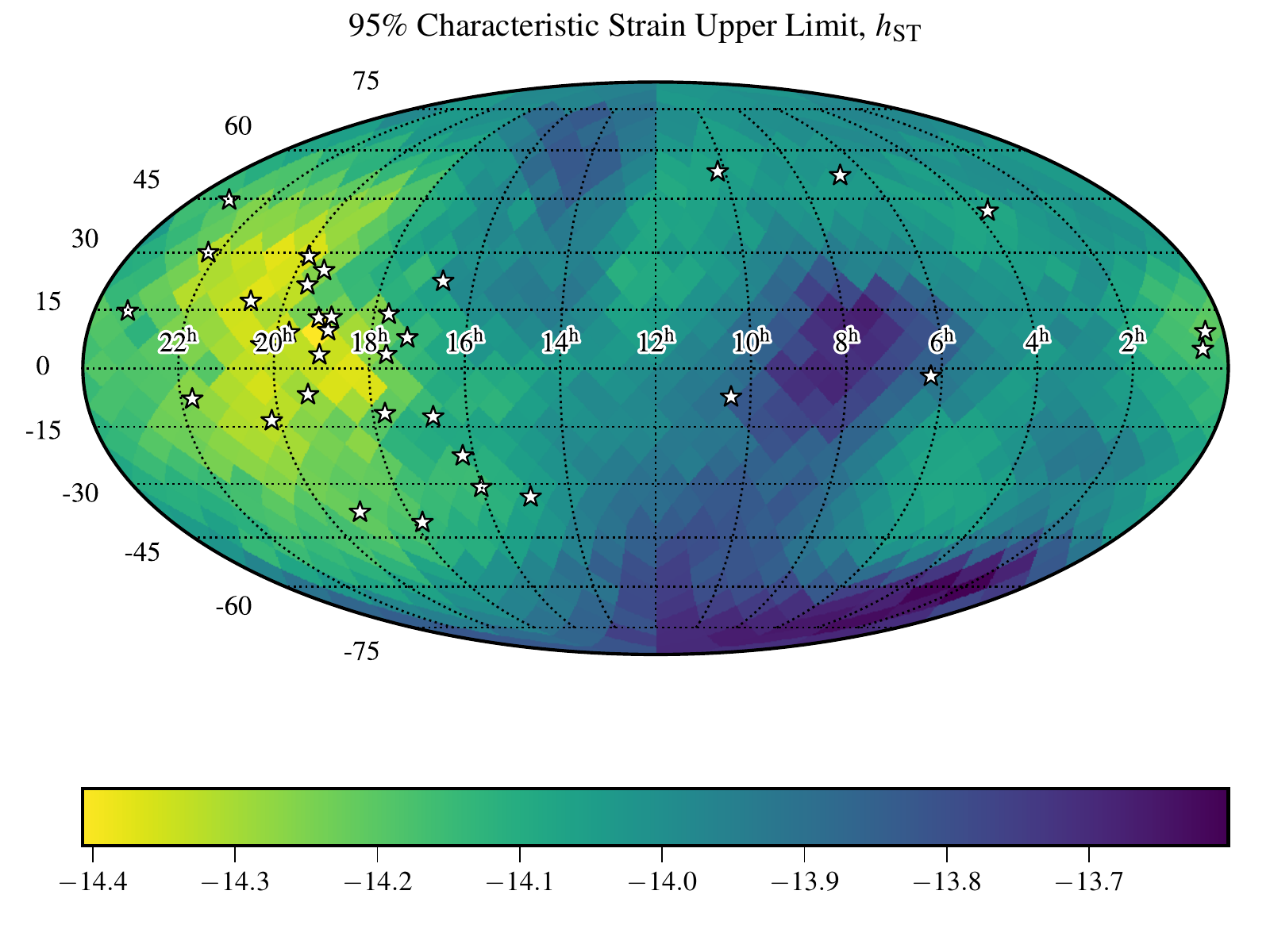}}\
	{\includegraphics[width = 3.25in]{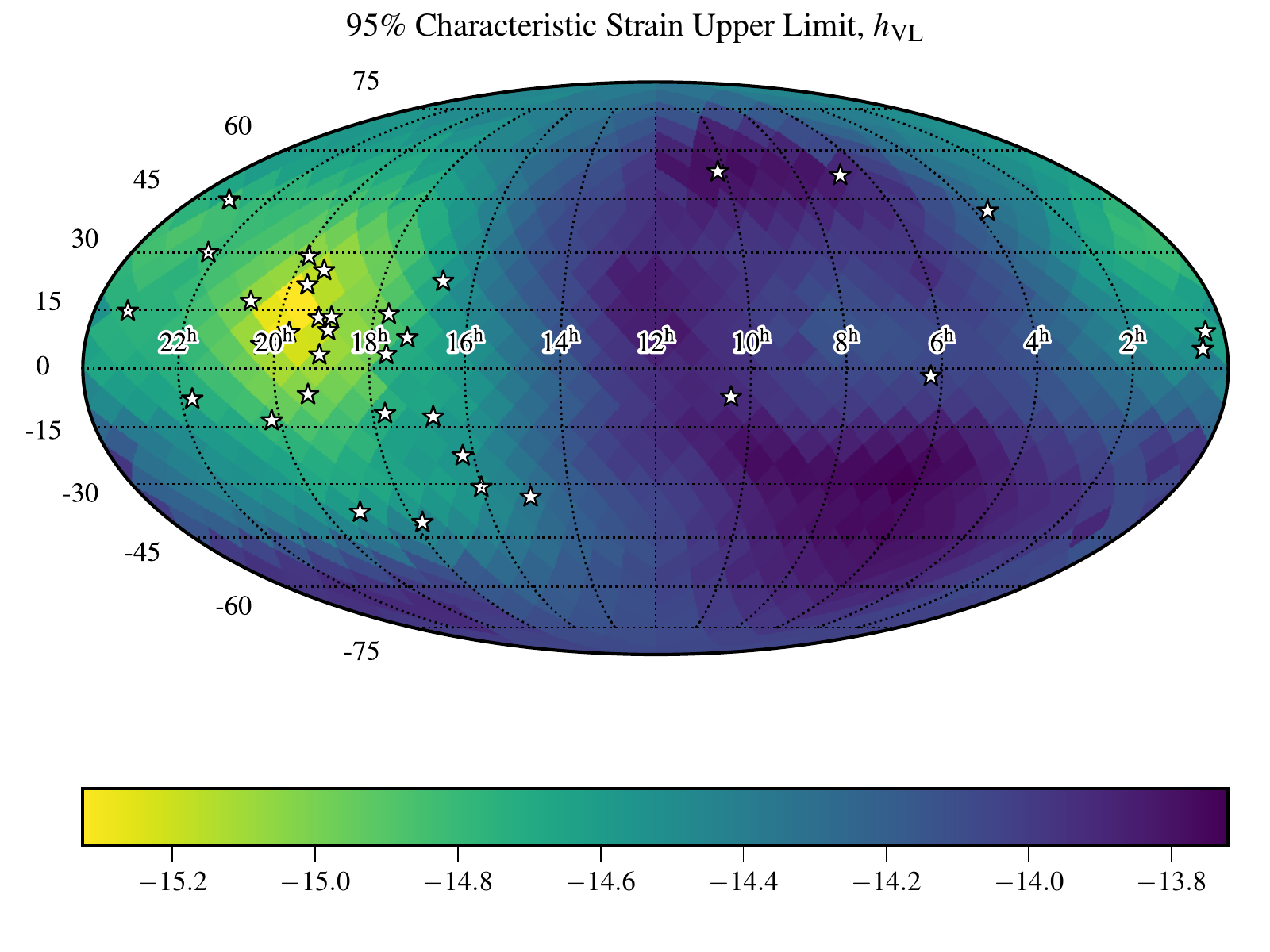}}
	{\includegraphics[width = 3.25in]{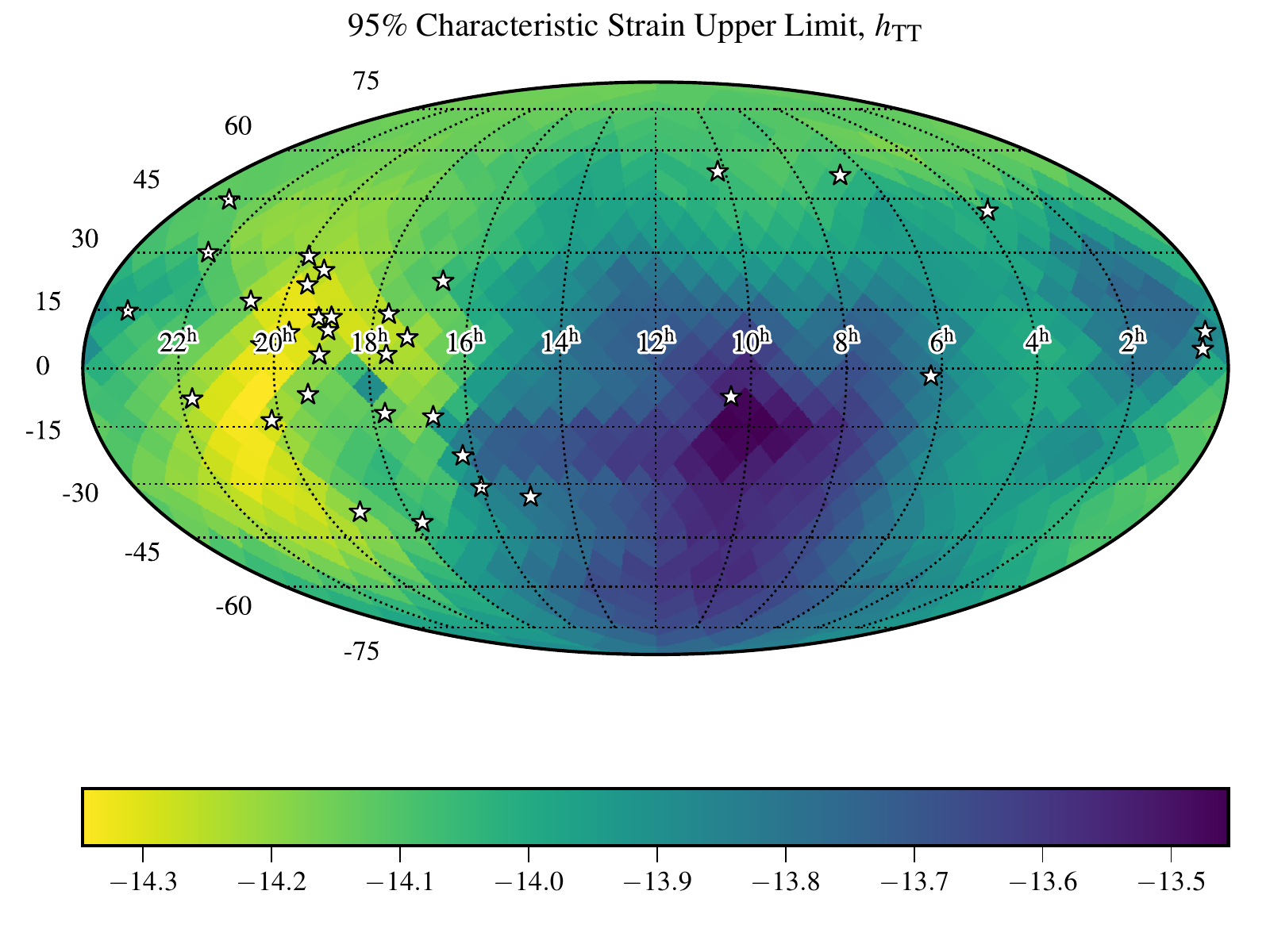}}
	\caption{Sky maps of an all modes upper limit search at fixed sky locations for GW frequency $f^{\text{TT}}_{GW} = 1\times10^{-8}$Hz. From left to right, top to bottom, the plots correspond to SL, ST, VL, and TT strains.}
	\label{fig:fixedsky}
\end{figure*}

\subsection*{Analysis 7: Upper Limits as a function of Sky Location}\label{sub7}
We already established the sky location bias of Analyses 3 and 4, as well as the independence of the upper limits when a TT signal is present in Analyses 5 and 6. It is, therefore, permissive to search for upper limits as a function of fixed sky location. This provides a more illuminating measure of the upper limit as a function of sky location when the upper limits span many orders of magnitude. In other words, we see in detail the array's sensitivity to localized alt-pol GWs based on the source's sky location. For a noise-only realization, the sky maps for marginalized strain upper limits at fixed sky locations are shown in Figure~\ref{fig:fixedsky}. We can see the enhanced sensitivity from longitudinal modes makes the upper limits more pronounced in sky regions with many more pulsars, namely the left side of the maps, as seen in the range of values indicated in the color bar. In the region of the sky that is well populated by pulsars, the upper limits on the amplitudes of the longitudinal modes are an order of magnitude lower than for the transverse modes. This is consistent with the enhanced response to longitudinal modes for sources near to the sky location of a pulsar seen in Figure~\ref{fig:antenna}.

\section{Conclusions}\label{conclusions}
We showed that for a modified gravity GW of total SNR$_{\text{eff}}\sim20$, we can detect the individual amplitudes of alt-pols even when their collective contribution to the SNR$_{\text{eff}}$ is $\sim10$, and the analysis is not hindered by a higher dimensional model. This is encouraging as this contribution is only moderately loud, and our array is medium-sized with 34 pulsars. If significantly quieter alt-pols exist relative to GR a priori, we would need to rely on a very loud TT component being present to detect them, and the enhanced response due to longitudinal modes can prove advantageous in this respect.

We put upper limits on alt-pols in the presence of a TT mode, whose detection is unaffected by the search over additional strains. The values of the alt-pol upper limits depend on the location of the TT mode source, and thus on the sky location in general. We tested this for two separate cases, with one signal originating close to J1024-0719, which has few pulsar neighbors, and the second originating behind the GC, near many pulsars. The latter case rendered smaller alt-pol upper limits due the enhanced longitudinal response. We showed that over a sky-averaged upper limit search in the absence of any signal, the posterior probability is diminished in regions of the sky where longitudinal modes are enhanced. We subsequently showed the increased range of the strain values of longitudinal modes in our upper limit sky maps with fixed source sky locations.

We conclude that the upper limits set on alt-pol strains will be independent of the presence of a TT signal and depend only on sky location. The upper limits will be meaningful if we detect a TT mode as it will allow us to place constraints on coupling constants of alternative theories of gravity relative to GR.

\section*{Acknowledgments}
We appreciate the support of the NSF Physics Frontiers Center Award PFC-1430284. We are grateful for computational resources provided by Leonard E. Parker Center for Gravitation, Cosmology and Astrophysics at the University of Wisconsin-Milwaukee, which are supported by NSF Grant 1626190. We thank Nicol\'{a}s Yunes for helpful discussions.
\bibliography{biblioNotes}

\end{document}